\documentclass[preprint2]{aastex}
\usepackage{natbib}

\newcommand{\arcsecpoint}{\ifmmode ''\!. \else $''\!.$\fi}

\newcommand{\kms}{\ifmmode {\rm km\ s}^{-1} \else km s$^{-1}$\fi}
\newcommand{\Msun}{\ifmmode {\rm M}_{\odot} \else M$_{\odot}$\fi}
\newcommand{\Lsun}{\ifmmode {\rm L}_{\odot} \else L$_{\odot}$\fi}
\newcommand{\qo}{\ifmmode q_{\rm o} \else $q_{\rm o}$\fi}
\newcommand{\Ho}{\ifmmode H_{\rm o} \else $H_{\rm o}$\fi}
\newcommand{\ho}{\ifmmode h_{\rm o} \else $h_{\rm o}$\fi}
\newcommand{\ltsim}{\raisebox{-.5ex}{$\;\stackrel{<}{\sim}\;$}}
\newcommand{\gtsim}{\raisebox{-.5ex}{$\;\stackrel{>}{\sim}\;$}}
\newcommand{\vFWHM}{\ifmmode v_{\mbox{\tiny FWHM}} \else
                    $v_{\mbox{\tiny FWHM}}$\fi}
\newcommand{\CCF}{\ifmmode F_{\it CCF} \else $F_{\it CCF}$\fi}
\newcommand{\ACF}{\ifmmode F_{\it ACF} \else $F_{\it ACF}$\fi}
\newcommand{\Halpha}{\ifmmode {\rm H}\alpha \else H$\alpha$\fi}
\newcommand{\Hbeta}{\ifmmode {\rm H}\beta \else H$\beta$\fi}
\newcommand{\Hgamma}{\ifmmode {\rm H}\gamma \else H$\gamma$\fi}
\newcommand{\Hdelta}{\ifmmode {\rm H}\delta \else H$\delta$\fi}
\newcommand{\Lya}{\ifmmode {\rm Ly}\alpha \else Ly$\alpha$\fi}
\newcommand{\Lyb}{\ifmmode {\rm Ly}\beta \else Ly$\beta$\fi}
\newcommand{\HeI}{\ifmmode {\rm He}\,{\sc i}\,\lambda5876 \else 
	          He\,{\sc i}\,$\lambda5876$\fi}
\newcommand{\HeII}{\ifmmode {\rm He}\,{\sc ii}\,\lambda4686 \else 
	           He\,{\sc ii}\,$\lambda4686$\fi}

\newcommand{\feii}{Fe\,{\sc ii}}
\newcommand{\feiii}{Fe\,{\sc iii}}

\newcommand{\ciii}{\ifmmode {\rm C}\,{\sc iii} \else C\,{\sc iii}\fi}
\newcommand{\civ}{\ifmmode {\rm C}\,{\sc iv} \else C\,{\sc iv}\fi}

\newcommand{\mgii}{Mg\,{\sc ii}}

\begin{document}

\slugcomment{to be published in Astrophys.Journal}

%
%

\title{\feii /\mgii\ Emission Line Ratio in High Redshift Quasars
{$^\ast$}\\ 
}

\author{M.\,Dietrich
\altaffilmark{1,2},
F.\,Hamann
\altaffilmark{1},
I.\,Appenzeller
\altaffilmark{3},
and
M.\,Vestergaard
\altaffilmark{4}
}
\altaffiltext{1}
{Department of Astronomy, University of Florida, 211 Bryant Space Science
 Center, Gainesville, FL 32611-2055, USA}   
\altaffiltext{2}
{Department of Physics \& Astronomy, Georgia State University, One Park Place
 South SE, Suite 700, Atlanta, GA 30303, USA}
\altaffiltext{3}
{Landessternwarte Heidelberg--K\"{o}nigstuhl, K\"{o}nigstuhl 12, 
 D--69117 Heidelberg, Germany}
\altaffiltext{4}
{Department of Astronomy, The Ohio State University, 140 West 18th Av.,
 Columbus, OH 43210-1173, USA\\[1mm]
 $^\ast$ Based on observations collected at the CTIO Observatory, Chile,
  at the European Southern Observatory, Paranal, Chile, and the the W.M.
  Keck Observatory, Hawai'i}

\email{dietrich@chara.gsu.edu}
%
%

\begin{abstract}
We present results of the analysis of near infrared spectroscopic observations 
of 6 high-redshift quasars ($z\ga 4$), emphasizing the measurement of the 
ultraviolet \feii/ \mgii\ emission line strength in order to estimate the 
beginning of intense star formation in the early universe.
To investigate the evolution of the \feii/\mgii\ ratio over a wider range in
cosmic time, we measured this ratio for composite quasar spectra which cover a 
redshift range of $0 \la z\la 5$ with nearly constant luminosity, as well as 
for those which span $\sim 6$ orders of magnitude in luminosity. 
A detailed comparison of the high-redshift quasar spectra with those of 
low-redshift quasars with comparable luminosity shows essentially the same 
\feii /\mgii\ emission ratios and very similar continuum and line spectral 
properties, i.e. a lack of evolution of the relative iron to magnesium 
abundance of the gas in bright quasars since $z\simeq 5$.
Current nucleosynthesis and stellar evolution models predict that 
$\alpha$-elements like magnesium are produced in massive stars ending in 
type II SNe, while iron is formed predominantly in SNe of type Ia with 
intermediate mass progenitors. This results in an iron enrichment delay
of $\sim 0.2 \,{\rm to}\,0.6$\,Gyr.
We conclude that intense star formation activity in the host galaxies of
$z\ga 4$ quasars must have started already at an epoch corresponding to 
$z_f \simeq 6 - 9$, when the age of the universe was $\sim 0.5$\,Gyr 
($H_o = 72$ km\,s$^{-1}$\,Mpc$^{-1}$, $\Omega _M = 0.3$, $\Omega _\Lambda = 
0.7$). This epoch corresponds well to the re-ionization era of the universe.
\end{abstract}

\section{Introduction}
There is growing evidence that galaxy formation and intense star formation are 
closely related to quasar activity, particularly in the early universe.
For example, the detections of large amounts of molecular gas and dust 
($M_{molec.}\simeq 10^{10} M_\odot$, $M_{dust}\simeq 5\times 10^8 M_\odot$) 
around quasars at redshifts $z\ga 4$ 
\citep{Isetal94,Ohetal96,Anetal99,Caetal00,Caetal02,Ometal96,Ometal01} 
establish a close relationship between quasars and massive galaxies.
There is also solid evidence that the co-moving space density of quasars and
the cosmic star formation rate both show a strong increase toward higher
redshifts \citep{Liletal96,Conetal97,TrMa98,Steetal99, Lanetal02}.
At low redshifts it has been found that characteristic physical properties of 
quasar host galaxies, e.g., bulge mass, luminosity and stellar velocity 
dispersion, are well correlated with the mass of the central black hole 
\citep{KoRi95,Gebetal00,MeFe01,FeMe01,Treetal02}.
Moreover, it has become clear that all massive galaxies today harbor super
massive black holes (SMBHs) in their centers, implying that at one time
(early in their evolution) these galaxies were hosts to active, i.e. quasar
or quasar-like, nuclei. 

Since quasars are among the most luminous objects in the universe they can be 
observed at practically any distance at which these objects occur. This
practical aspect together with their close relation to galaxy and star
formation history makes quasars excellent probes to study these processes
in the early universe.
Assuming that the gas of the broad emission line region (BELR) originates from
the interstellar medium of the quasar host galaxies, these lines provide 
valuable information on the chemical composition and enrichment history of the 
quasar host galaxies.
Analysis of ultraviolet broad emission lines (BELs), as well as intrinsic 
associated absorption lines in quasar spectra, indicate that the gas-phase
metallicities near quasars are at least solar to several times solar at even
the highest measured redshifts
\citep[e.g.,][]{HaFe92,HaFe93,HaFe99,Feetal96,Koetal96,Hama97,Dietal99,pet99,
DiWi00,Waetal02,Dietal03a,Dietal03b}.
These results show that well before the epoch corresponding to 
$z\simeq 3\,{\rm to}\,4$ significant star formation must have taken place in 
the galactic or proto-galactic cores where these quasars reside 
\citep[e.g.,][]{CeOs99} to enrich the gas phase.

The epoch of first intense star formation activity in quasar host galaxies can 
also be dated exploring the abundance of $\alpha $-process elements relative to
iron. According to present chemical enrichment scenarios, $\alpha $-element 
nuclei are produced predominantly in type II SNe with massive progenitors on 
time scales of $\tau _{evol}\simeq 2 - 10$ Myr.
On the other hand, the dominant source of iron is assumed to be type Ia SNe at
the end point of the evolution of intermediate mass stars in binary systems, 
about $\tau _{evol}\simeq 1$\,Gyr after the onset of the star formation epoch
\citep[e.g.,][]{Tins79,MaGr86,WST89,Yoetal96}.
The amount of iron returned to the interstellar medium in SN\,II ejecta is 
rather low \citep{Yoetal96,Yoetal98}. 
It is generally assumed that the significantly different time scales of the 
release of $\alpha$-elements and iron to the interstellar medium results in a 
time delay of the order of $\sim 1$\,Gyr in the iron enrichment.
However, recent studies by \citet{Matt94}, \citet{FrTe98}, and \citet{MaRe01} 
indicate that this delay may be $\sim 0.2 - 0.6$\,Gyr for elliptical galaxies.
Detecting \feii\ emission at high redshift comparable to the relative strength
observed in quasars at lower redshift indicates that the formation of the stars
which had released the iron had occurred $\sim 0.3 - 0.8$\,Gyr earlier in the
quasar host galaxy.
Therefore, the line ratio of $\alpha $-element vs. iron emission can be used 
as a cosmological clock \citep[e.g.,][]{HaFe99,MaRe01}.

A suitable spectral feature for representing the presence of $\alpha$-elements 
is the prominent \mgii\ resonance doublet at $\lambda \lambda 2795, 2803$\,\AA\
(hereafter \mgii $\lambda 2798$). 
The generally strong UV \feii\ multiplets around \mgii\ in quasar spectra 
\citep[$\lambda \lambda \sim 2000 - 3000$\,\AA ;][]{Wietal80,Gran81,Neetal85}, 
are probably the most promising for obtaining at least some information on the
relative iron abundance \citep[e.g.,][]{WaOk67,WNW85,HaFe99}.
Since Mg$^+$ and Fe$^+$ have similar ionization potentials (7.6\,eV and 
7.9\,eV, respectively), it is generally assumed that \mgii $\lambda 2798$ and 
the \feii\ emission both originate in the partially ionized zone of the 
line emitting gas. 
However, it is important to keep in mind that the \mgii $\lambda 2798$ and 
\feii\ emission are differently affected by the conditions of the line 
emitting gas.
As \citet{WNW85} pointed out, the Fe$^+$ zone is much more extended than Mg$^+$
zone and contains hot regions where magnesium is ionized to Mg$^{++}$, which
emits \mgii\ much less efficiently.
Furthermore, \mgii $\lambda 2798$ and the \feii\ emission show significantly 
different radiative transfer properties.
While \mgii $\lambda 2798$ is a pair of two resonance lines \feii\ involves 
several ten thousand lines. Due to these thousands of transitions, \feii\ is
subject to processes like line fluorescence, Ly$\alpha $ pumping and turbulence
\citep[e.g.,][]{Veetal99} which results in a complicated line transfer. 
It is also known that UV\,\feii\ emission suffers destruction by Balmer 
continuum absorption and conversion to optical \feii\ emission for large
column densities, $N_H\gtsim 10^{24}$\,cm$^{-2}$, \citep{WNW85,Joly87}.
Since \mgii $\lambda 2798$ forms predominantly in regions of higher Balmer 
continuum opacity and these lines are closer to the wavelength of the Balmer 
edge than the strong UV \feii\ emssion, \mgii $\lambda 2798$ emission is more 
reduced than UV \feii\ by Balmer continuum absorption \citep{WNW85}. 
Another process which affects the \feii\ and \mgii $\lambda 2798$ emission are 
charge exchange reactions. As pointed out by \citet{Joly87} for
$T_e \gtsim 10^4$\,K Fe$^+$ is ionized to Fe$^{++}$ by charge exchange 
reactions while Mg$^+$ is nearly unaffected by this process.
Given the complexity of the \feii\ emission spectrum accurate iron abundances 
are not yet straight forward to deduce and a full synthesis of individual 
quasar spectra is required. This seriously complicates such measurements 
because the \feii\ emission is hard to synthesize 
\citep[e.g.,][]{Veetal99,CoJo00,SiPr98,SiPr03}.

However, in spite of these uncertainties which affect the \feii\ and the 
\mgii $\lambda 2798$ line emission, the \feii /\mgii\ emission ratio also must
reflect the Fe/Mg abundance. While, in view of the uncertainties concerning the
detailed physical processes, it may not be possible to derive absolute 
abundances from these lines, it should, nevertheless, be possible to derive at
least variations of th Fe/Mg ratio with redshift from emission line ratios if 
objects with similar observed physical properties are compared. 
This is feasible owing to the similar appearance of their optical-UV spectra 
across the known redshift and luminosity ranges \citep{Dietal02b} and the 
observed similar variability properties \citep{PeWa99,PeWa00,OnPe02},
suggesting a generally similar physical structure of the AGN emission-line 
regions.
Assuming that the physical conditions of the line emitting gas are comparable 
for different objects, the effects mentioned above should influence the 
\feii /\mgii $\lambda 2798$ line ratio measured in individual quasars in a very
similar way. To minimize the uncertainties that are introduced by, for example,
different radiative transfer properties and different conditions of the line 
emitting gas ($T_e, N_H, n_e$), we concentrate 
(a) on individual quasars of similar intrinsic luminosity and 
(b) on composite quasar spectra of comparable luminosity as the individual high
    redshift quasars.
This approach, adopted in this work, reduces the influence of a possibly 
luminosity dependent spectral energy distribution (SED) of the incident 
ionizing continuum on the physical conditions of the line emitting region. 
As a consequence, similar continua will result in similar conditions of the 
line emitting gas, emphasizing effects of relative abundance variations.
Assuming that the physical parameters affecting \feii/\mgii\ are roughly the 
same at all redshifts, then the average of measured \feii/\mgii\ ratios at 
different redshifts will yield useful constraints on the redshift evolution 
of Fe/Mg.

Observations of the \mgii $\lambda 2798$ and of the rest-frame UV-multiplets 
of \feii\ in the spectra of several high-redshift quasars have been published 
by \citet{Hietal93}, \citet{Eletal94}, \citet{Kaetal96}, \citet{Taetal97}, 
\citet{Muetal98,Muetal99}, \citet{Yoetal98}, \citet{Thetal99}, 
\citet{Gretal01}, \citet{Dietal02a}, \citet{Iwetal02}, and \citet{Freu03}.
Generally, fairly strong \feii\ emission was found, indicating a very early 
cosmic epoch of the first star formation in the high-redshift quasar host 
galaxies. 

In order to obtain more quantitative information on the beginning of the first 
star formation episode in quasar host galaxies at high redshifts, we extend our
study of 6 high-redshift quasars $z \simeq 3.4$ \citep{Dietal02a} to earlier 
cosmic epochs. We observed 6 quasars with redshifts $4.4 \la z \la 5$ to extend
the redshift range and increase the sample size.
This high redshift range corresponds a time when the universe was less than 
$\sim 1.3$\,Gyr old 
\citep[$H_o  = 72$ km\,s$^{-1}$ Mpc$^{-1}$, $\Omega _M = 0.3$, 
$\Omega _\Lambda = 0.7$;][]{CPT92,Freetal01}.
To put this study on the evolution of the \feii/\mgii\ ratio at high redshifts
in a wider cosmological context, we analyzed quasar composite spectra which we 
calculated for a wide range in luminosity ($\sim 6$ orders of magnitude) and 
redshift ($0 \la z \la 5$) to study continuum and emission line correlations,
i.e. the Baldwin-Effect \citep{Dietal02b}. 

We find a lack of evolution of the measured \feii /\mgii\ emission line ratios
in these high-redshift quasars compared to those in quasar spectra at 
low-redshifts.
This indicates that the relative iron to magnesium abundance of the gas in 
bright quasars is unchanged since $z\simeq 5$.
We conclude that intense star formation activity in the host galaxies of the
studied high-redshift quasars must have started already at an epoch 
corresponding to $z_f \simeq 6 - 9$, when the age of the universe was 
$\sim 0.5$\,Gyr. This epoch corresponds well to the re-ionization era of the 
universe.

\section{Observations and Data Analysis}
We observed 6 quasars with redshifts $4.4\la z \la 5.0$ to measure the 
\feii /\mgii\ ratio in the restframe ultraviolet (hereafter referred to as 
the $z=4.5$ sample). The spectra were recorded
using infrared spectrographs of the W.M.\,Keck Observatory at Hawai'i, the 
Cerro Tololo Inter-American Observatory (CTIO) at Chile, and the ESO VLT 
Paranal Observatory at Chile (Table 1). 
The quasars were selected to be accessible at the time of observation and
to be bright enough to be observed within reasonable integration times.
Furthermore, UV emission lines like \civ $\lambda 1549$ are uncontaminated by 
strong and broad absorption features.
For the selected redshifts the prominent \mgii $\lambda 2798$ emission line 
and most of the strong ultraviolet \feii\ emission at $\lambda \lambda 2300 - 
2600$\,\AA\ are redshifted into the near infrared H-band. We covered 
practically the whole restframe wavelength range 
$\lambda \lambda  2200 - 4500$\AA\ continuously, which made it possible to 
derive reliable continuum fits and, thus, at least internally accurate flux 
values for the ultraviolet  \feii\ emission. 
In addition to the $z\ga 4.4$ quasars, Q\,2050$-$358 ($z=3.5$) was recorded to 
enlarge our sample of quasars at $z\simeq 3.4$ (hereafter `the $z=3.4$ 
sample').
The details of the data aquisition and processing are described in the
following subsections.

\subsection{Keck\,II NIRSPEC Observations}
The high redshift quasar PC\,1158$+$4635 \citep[z=4.733;][]{Schetal89} was 
observed in service mode with NIRSPEC at KECK\,II in the near infrared domain 
on May 19/20, 2000.
The seeing was $\sim 0\arcsecpoint 5$, estimated from the spatial profile of 
the standard star spectra.
The 1024 $\times$ 1024 InSb detector array was used in the longslit mode 
(0\arcsecpoint 76$\times$ 42\arcsec , 0\arcsecpoint 19/pixel).
To cover the near infrared spectral range three settings of the grating were 
used which yield the wavelength ranges $(1.26 - 1.54)\mu$m, 
$(1.52 - 1.81)\mu$m, and $(2.05 - 2.48)\mu$m which cover the traditional
J, H, and K bands, respectively.
Making use of the slit length the location of the object along the slit was 
alternated by $\sim 20$\arcsec\ for subsequent exposures to assure a proper 
sky-correction.
Due to the high brightness of the night sky in the near infrared, the 
integration times of the individual exposures of the quasar were limited 
to 120\,sec and 400\,sec, respectively. 
The observation log and the total integration times for the individual bands
are given in Table 1.

A mean dark frame was calculated using 5 individual dark frames.
Flatfield frames were taken in the lamp-on and lamp-off mode for each
wavelength range. 
The frames corrected for thermal background were used to compute normalized 
flatfield frames for the flatfield correction \citep{Horn86}.
Argon comparison spectra were recorded for wavelength calibration for each 
wavelength range.
In the J to H infrared wavelength range the spectral resolution, measured using
the FWHM 
of strong night sky emission lines, amounts to R\,=\,570 corresponding to 
$\Delta v \simeq 520$ km\,s$^{-1}$ and in the K wavelength range it amounts 
to R\,=\,650 ($\Delta v \simeq 460$ km\,s$^{-1}$).
Because the sky brightness can vary significantly on time scales of minutes
(comparable to our exposure times) we used the sky spectrum measured along
the slit in each frame to subtract the sky emission.
A $3^{rd}$ to $6^{th}$ order polynomial fit was calculated for each 
wavelength element to describe the spatial intensity 
distribution of the night sky emission. 
The fit based on $\sim 18\arcsec $ wide regions which were $\sim 5\arcsec $
separated from quasar or standard star.

The standard star HD\,105601 \citep[$m_K = 6.685$, A2V;][]{Eletal82,Arnetal89} 
was observed for each setting before or after the observations of the quasar 
spectrum for flux calibration and correction of the strong atmospheric 
absorption features. 
To obtain a sensitivity function we assumed that the spectral energy 
distribution of HD\,105601 can be described in the near infrared wavelength 
range with a black body spectrum. We applied $T_{eff} = 8995$\,K to computed a 
black body spectrum for the observed wavelength range.
This black body spectrum was scaled to match the apparent magnitudes of 
HD\,105601 in the J-,\,H-, and K-band \citep{Eletal82}.

We used the spatial curvature of the standard star spectrum to trace the 
location of the slightly curved quasar spectrum. The spectra of the individual
exposures were extracted using an optimal extraction routine \citep{Horn86}. 

The 1\,D quasar spectra were corrected for broad atmospheric absorption bands 
using an appropriately scaled transmission function derived from HD\,105601. 
Cosmic-ray events in the individual 1\,D spectra were corrected by comparing 
the spectra with each other.
Finally, we computed a weighted mean of the quasar spectra for each setting,
with the weight given by the signal-to-noise ratio of the continuum. 
These mean quasar spectra were flux calibrated using the corresponding 
sensitivity function for the wavelength range. The mean
spectra of each setting were rebinned to a uniform stepsize in wavelength and
merged. 
The flux calibrated spectrum of PC\,1158+4635 is shown in Figure 1. 

\subsection{CTIO Observations}
On Sept.\,14 to 16, 2000 we used the near infrared spectrometer and camera 
OSIRIS attached to the 4\,m Blanco telescope at Cerro Tololo Inter-American 
Observatory (CTIO), Chile, to observed BRI\,0019-1522, BR\,0103+0032, 
PSS\,J0248+1802, Q\,2050$-$359, and BRI\,2237-0607. 
The seeing was typically $\sim 0\arcsecpoint 8$ during these three nights. 
The detector was a $1024 \times 1024$ Hawai'i HgCdTe array ($18.5\mu$m/pixel). 
We used OSIRIS in the low-resolution cross-dispersed mode 
with a longslit of 1\arcsecpoint 2 $\times$ 30\arcsec. The pixel scale was 
0\arcsecpoint403/pixel. 
In the cross-dispersed mode the J, H, K bands are recorded in adjacent orders 
in a single exposure yielding the entire near infrared wavelength range, i.e.,
$(\sim 0.9$ -- $2.5)\mu$m. 
Making use of the long-slit mode we placed the object at seven different
locations along the slit, separated by 3\arcsec\ each for subsequent 
exposures to optimize the sky-correction.
For flux calibration and correction of the strong atmospheric absorption 
bands, separating the J- and H-band and the H- and K-band, respectively, and 
additional weaker atmospheric (absorption) features we observed standard stars 
several times during each night \citep{Eletal82,Peetal97}.
In Table 1 the exposure times for the quasars are given.

For each of the different individual exposure times of the science frames, 
i.e., quasars (200\,sec, 300\,sec) and standard stars (3\,sec to 20\,sec), we 
recorded 20 to 25 dark frames. They were used to calculate average dark frames 
for each specific exposure time. The quasar and standard star spectra were dark
corrected, subtracting the average dark frame of the corresponding exposure
time.

Flatfield frames were taken in the lamp-on and lamp-off mode, both with 200
and 300 seconds integration time, respectively. The average lamp-off mode
frames were subtracted from the average lamp-on mode frames to correct for the
thermal background.
These thermal background corrected mean flatfield frames were split up into 
separate 2\,D spectra for the J, H, and K band spectrum. Those were used to 
compute normalized flatfield frames \citep{Horn86} which we employed for 
2\,D flatfield correction for each individual spectral band.

For wavelength calibration we took helium-argon-neon comparison spectra.
The location of the emission lines in the individual calibration frames
differ by less than $\ltsim 0.1$ pixel compared to each other. Therefore, an 
average calibration frame was computed and was used for the wavelength 
calibration.
The uncertainty of the wavelength calibration is $\Delta \lambda
\sim 0.3$\,\AA . The spectral overlap of the individual spectral bands is of
the order of $\sim 350$\,\AA\ in the quasars restframe which allow a solid 
intercalibration to obtain a continuous spectrum of the near infrared 
wavelength range.
The achieved spectral resolution amounts to $R\simeq 1300$ in J, H, and K-band,
respectively, based on the measured FWHM of strong night sky emission lines and
isolated lines of the wavelength calibration frames.

We corrected each frame individually for the night sky intensity. For this 
purpose a $3^{rd}$ order polynomial fit was calculated for each wavelength 
element to fit the spatial intensity distribution of the night sky emission. 
The fit was based on two regions, $2\arcsec $ to $20\arcsec $ wide, depending 
on the location of the object along the slit. These regions were separated by 
$\sim 10\arcsec $.

To derive a sensitivity function we assumed that the IR spectral energy 
distribution of a A-type and G-type star can be described with a black 
body spectrum of a specific $T_{eff}$. 
We used the black body temperatures given by \citet{Kuru92} for the 
different spectral types of the stars we observed. 
We applied $T_{eff} = 5850$\,K (HD\,1205, HD\,25402, HD\,198678), 
           $T_{eff} = 8250$\,K (HD\,19904, HD\,205772), 
       and $T_{eff} = 8790$\,K (HD\,2811) to compute a 
black body spectrum for the observed wavelength range.
The black body spectra were scaled to match the apparent magnitudes of the 
observed standard stars in the J-,\,H-, and K-band.
For HD\,2811, HD\,19904, and HD\,205772 we used the J, H, and K magnitudes 
given by \citet{Eletal82}, while for HD\,1205, HD\,25402, and HD\,198678 we 
had to calculate the apparent near infrared magnitudes based on the apparent 
V-band magnitude.
A sensitivity curve was calculated for each standard star.
In general, the sensitivity functions given by the individual stars are 
nearly identical in shape and strength, within less than 2\,\% . However, 
for a $\sim 200$\,\AA\ wide region at the beginning and the end of each 
spectral wavelength range they differ less than $\sim 20$\% . But these 
regions fall into the strong absorption bands which separate the J, H, and 
K-band. We used the individual sensitivity functions obtained from the 
standard stars, we observed, to compute a mean sensitivity function for each 
wavelength range.

The quasar spectra were extracted using an optimal extraction algorithm
\citep{Horn86}. The width of the spatial profile for the quasar spectra was 
the same as measured for the stars because high-redshift quasars can be treated
as point sources. The individual spectra of the quasars were flux calibrated 
employing the mean sensitivity function of the corresponding spectral band.
To correct for cosmic-ray events the individual 1\,D spectra were compared 
with each other. A weighted mean spectrum was calculated for each quasar.
The weight was given
by the mean signal-to-noise ratio in the continuum across the spectrum.
The mean quasar spectra were corrected for atmospheric absorption using 
appropriately scaled transmission functions provided by observed spectra of 
the standard stars.
The flux calibrated quasar spectra observed with OSIRIS at the 4\,m Blanco 
telescope are shown in Figure 1.

\subsection{Paranal VLT/ISAAC Observation}
The high-redshift quasar SDSS\,1204$-$0021 \citep[$z = 5.03$;][]{Fanetal00} was
observed in service mode with ISAAC attached to the VLT\,UT1 {\it Antu} in the
H-band spectral range. 
The seeing was estimated from the spatial profile of standard star spectra 
to $\sim 1\arcsecpoint 1$ (April 11, 2001) and $\sim 1\arcsecpoint 2$ 
(June 14, 2001).
A Hawai'i $1024 \times 1024$ array from Rockwell was used in the longslit 
mode ($1\arcsec \times 120 \arcsec$, $0\arcsecpoint 147/$pixel).
We varied the location of the object along the slit by $60\arcsec $ (with an 
additional random offset within $5\arcsec $ for subsequent exposures to 
optimize the sky correction).
For flux calibration and correction of the strong atmospheric absorption bands
at the beginning and end of the H-band spectrum a standard star was 
observed each night.
On April 10 and June 14, 2001, respectively, ten quasar spectra with 180 sec 
exposure time were recorded which results in a total exposure time of 1 hr
(Table 1). 

For each night three dark frames with 5 sec (mean of 10 individual exposures)
and 180 sec exposure time (mean of 3 individual exposures) were recorded. 
These frames were averaged for the corresponding exposure time and used for 
dark correction.
Flatfield frames were taken in the lamp-on and lamp-off mode for each night. 
The lamp-off mode frames were subtracted from the adjacent lamp-on mode frame.
The resulting flatfield frames were averaged and were normalized to apply a
2\,D flatfield correction.

Argon-xenon comparison spectra were recorded for wavelength calibration. 
The wavelength range amounts to $\sim 14175$ to $19040$\,\AA\ (April 10)
and $\sim 14195$ to $19065$\,\AA\ (June 14) with a stepsize of 
$\sim 4.8 \pm 0.5$\,\AA /pixel. The wavelength calibration is based on 20 
individual lines. The spectral resolution, measured using the FWHM of strong 
night sky emission lines and isolated lines in the argon-xenon spectra, amounts
to $R \simeq 570$.

We subtracted the night sky intensity for each spectrum individually.
A $3^{rd}$ order polynomial fit was calculated for each wavelength element 
to describe the spatial intensity distribution of the night sky emission. 
The fit based on two regions, $\sim 25\arcsec $ and $\sim 80\arcsec $ wide,
which were separated by $\sim 8\arcsec $. 

To derive a sensitivity function the standard stars HD\,87097 (April 10) and
HD\,143567 (June 14) were observed.
We used the black body temperatures given by \citet{Kuru92} with 
$T_{eff} = 5850$\,K (HD\,87097, G3V) and 
$T_{eff} = 10700$\,K (HD\,143567, B9V) 
to compute a black body spectrum for the H-band range.
These black body spectra were scaled to match the apparent magnitudes of the 
observed standard stars. A sensitivity curve was calculated for each standard 
star based on the V-band magnitude.

The spectra of the quasars which can be regarded as point sources, were 
extracted using an optimal extraction algorithm. The width of the spatial 
profile for the quasar spectra was the same as measured for the standard stars.
 
The individual spectra of SDSS\,1204$-$0021 were flux calibrated employing the
corresponding sensitivity function.
To correct for cosmic-ray events the individual 1\,D spectra were compared 
with each other.
For each quasar a weighted mean spectrum was calculated. The weight was given
by the mean signal-to-noise ratio in the continuum across the spectrum.
The mean quasar spectra were corrected for atmospheric absorption using 
appropriately scaled transmission functions provided by observed spectra of 
the standard stars.
Finally, the spectra obtained in April and June 2001 were averaged according 
to the signal-to-noise ratio in the continuum.
In Figure 2 the flux calibrated H-band spectrum of SDSS\,1204$-$0021 is shown.

\subsection{Composite Quasar Spectra}

The analysis of composite spectra has the advantage of more clearly 
representing the average properties of the sample used to calculate each 
composite. 
We recently compiled a large sample of rest-frame ultraviolet and optical 
spectra for more than 800 type 1 AGNs \citep{Dietal02b}.
About $\sim 300$ of these spectra cover the ultraviolet \feii\ -- \mgii\ range 
($\lambda \lambda 2000 - 3000$\,\AA ). 
A large fraction of the spectra were obtained by several groups for different 
studies over the last 20 years using ground-based instruments as well as 
{\it International Ultraviolet Explorer (IUE)} and {\it Hubble Space Telescope
(HST)}. In addition, we retrieved all publicly available AGN spectra from the 
{\it HST} MAST archive.
The type 1 AGNs (i.e., Seyfert\,1 and quasars) of the sample cover a redshift 
range of $0\la z \la 5$ and nearly 6 orders of magnitude in luminosity. 
We excluded Broad-Absorption Line quasars (BAL\,QSOs) when calculating the
composite spectra, although there are indications that BAL\,QSO emission line 
properties do not differ from non-BAL quasars \citep{Weyetal91}. 

We analyzed these composite spectra which we calculated to investigate 
continuum vs. emission line relations \citep{Dietal02b} for the \feii/\mgii\ 
ratio. We selected a narrow luminosity range to calculate composite spectra for
different redshift bins with nearly constant luminosity (the redshift sample). 
The location of this luminosity range was chosen to achieve a comparable
number of individual quasars contributing to each redshift bin in the
$z$ -- $\lambda L_\lambda (1450{\rm \AA })$ plane. 
We used a luminosity range of 
$46.16 \leq \log\,\lambda L_\lambda (1450{\rm \AA }) \leq 47.16$ erg\,s$^{-1}$
(i.e., $43.0 \leq \log\,L_\lambda (1450{\rm \AA }) \leq 44.0$).
This range contains 323 quasars with $z\geq 0.5$ and an average luminosity of
log\,$\lambda L_\lambda (1450{\rm \AA }) = 46.67\pm 0.12$ erg\,s$^{-1}$. 
We computed composite spectra for redshift bins with $\Delta z = 0.5$
centered on redshift $z = 0.75,1.25,...,4.25, {\rm and}\, 4.75$, 
respectively \citep{Dietal02b}.
The composite spectra characterized by nearly constant luminosity provide
valuable \feii/\mgii\ measurements as a function of redshift.
The quasar sample was also divided into 11 subsets with respect to the 
luminosity $L_\lambda (1450{\rm \AA})$ (the luminosity sample, see Table 2). 
Except for the lowest luminosity bin, each composite spectrum covers a range 
of $\Delta \log\,\lambda L_\lambda = 0.5$ in luminosity starting at 
log\,$\lambda\,L_\lambda (1450{\rm \AA }) = 43.16$ erg\,s$^{-1}$.
The composite spectra with increasing luminosity yield information on the
\feii/\mgii\ ratio as a function of luminosity.

\section{Fitting of the Quasar Spectra}

To measure the \feii\ and \mgii\ emission strengths, we first transformed the 
individual high-redshift quasar spectra into the restframe.
Although the \mgii $\lambda 2798$ emission line doublet sometimes appears to be
relatively isolated in quasar spectra, the superposition of many thousands of
discrete ultraviolet \feii\ emission lines forms broad emission blends 
\citep[e.g., Wills et al. 1985, hereafter WNW85;][]{Veetal99,SiPr98,SiPr03},
which tend to blend with the broad base of \mgii\ complicating the measurement
of \mgii\ and \feii .
Because of the line blending and the related difficulty defining the continuum 
level, these lines cannot be measured individually.
However, as demonstrated by WNW85, it is possible to obtain the
\feii\ emission strength in a multi-component fit approach. 
Therefore, we assumed our observed quasar spectra (z=3.4 and z=4.5 sample), as
well as the composite quasar spectra to consist of a superposition of 
the following four components, 
    (i)   a power law continuum (F$_\nu \sim \nu ^{\alpha}$),
    (ii)  Balmer continuum emission,
    (iii) a pseudo-continuum due to merging \feii\ emission blends,
and (iv)  an emission spectrum of other individual broad emission lines, such
as \mgii .
The power law continuum fit to the individual and composite quasar spectra was 
iteratively determined taking into account the measured strength of the \feii\ 
emission and Balmer continuum emission derived by the fitting process.
We already applied this multi-component fit approach successfully to a small 
sample of high redshift quasars at $z\simeq 3.4$ \citep{Dietal02a}.

\subsection{Non-stellar Continuum}
We first determined the underlying non-stellar power law continuum, component 
(i), from spectral windows which are free (or almost free) of contributions by
the components (ii) to (iv). For this purpose it is very important that the 
spectrum covers a wide wavelength range. The high-redshift quasars we observed 
allow us to use a 20\,\AA\ wide continuum window, centered at 
$\lambda _c \simeq 4000 \pm 50$\,\AA\ which is nearly free of emission line 
contamination and hydrogen Balmer continua. 
Only minor \feii -emission can be expected for this wavelength region 
\citep[WNW85;][]{Veetal99,CoBo96} and hence will not significantly affect the
continuum setting.
For 4 of the individual high-redshift quasars (BRI\,0019-1522, BR\,0103+0032, 
PC\,1158+4635, BRI\,2237-0607) we have also restframe UV spectra. These spectra
provide continuum windows at
$\lambda \lambda \simeq 1330 - 1380$\,\AA, and 
$\lambda \lambda \simeq 1440 - 1470$\,\AA, and
$\lambda \lambda \simeq 1685 - 1695$\,\AA\ which are nearly uncontaminated by 
line emission as well \citep{Fraetal91}.
For those quasars for which we had no access to the short wavelength UV part of
the spectrum (PSS\,J0248+1802, Q\,2050$-$358) we used continuum windows at 
$\sim 2230$\,\AA\ to estimate the continuum strength. The UV data allow us even
better constraints on the 
non-stellar continuum strength for these objects. The uncertainty introduced 
by estimating the continuum level, based on at least two of the spectral 
ranges described above, is estimated to be of the order of less than
$\sim 10$\,\% .

\subsection{Balmer Continuum}
To estimate the Balmer continuum emission spectrum for our fits we assumed gas 
clouds of uniform temperature ($T_e = 15000$\,K) which are partially optically 
thick. In this case the Balmer continuum spectrum ($\lambda \la 3646$\,\AA )
can be described by

$$ F_\lambda ^{BaC} = F^{BE} B_\lambda (T_e) (1 - e^{-\tau _\lambda}); 
   \quad \lambda \leq \lambda_{BE}$$

\noindent
with $B_\lambda (T_e)$ as the Planck function at the electron temperature
$T_e$, $\tau _\lambda$ as optical depth at $\lambda$, and $F^{BE}$ as a 
normalized estimate for the Balmer continuum flux density at the Balmer edge 
at $\lambda = 3646$\,\AA\ \citep{Gran82}.
The strength of the Balmer continuum emission can be estimated from the flux 
density at $\lambda \simeq 3675$\,\AA, after subtraction of the power-law 
continuum component, since at this wavelength there is no significant 
contamination by \feii\ emission \citep[WNW85;][]{Veetal99}.  
The $\lambda 3675$\,\AA\ restframe flux density level was therefore used to 
normalize the Balmer continuum spectrum.
At wavelengths $\lambda \geq 3646$\,\AA\ higher order Balmer lines are merging 
to a pseudo-continuum, yielding a smooth rise to the Balmer edge (WNW85).
We used the results of the model calculations provided by \citet{StHu95} 
(case B, $T_e = 15000$\,K, $n_e = 10^8 - 10^{10}$\,cm$^{-3}$) to estimate the
strength of high order Balmer emission lines, assuming a Gaussian profile for 
each high order Balmer line (FWHM = 3000 km\,s$^{-1}$).
We calculated several Balmer continuum spectra for $T_e = 15000$\,K and
$0.1 \leq \tau _\lambda \leq 2$ to obtain Balmer continuum emission templates.
These Balmer continuum templates were supplemented for $\lambda > 3646$\,\AA\ 
with high order Balmer emission lines with $10 \leq n \leq 50$, i.e., 
H$\vartheta $ and higher.

\subsection{\feii\ Emission}
Calculating the \feii\ emission spectrum is much more difficult and the 
influence of unknown parameters such as internal turbulence velocities, 
emission line transport, pumping by the incident continua, line fluorescence, 
and metallicity, which affect the emergent \feii\ spectrum,
are still not well understood 
\citep[e.g., WNW85;][]{Neetal85,Joly87,BaPr98,SiPr98,Veetal99,CoJo00}. 
In spite of these uncertainties it has been shown that the \feii\ emission 
strength of Seyfert\,1 galaxies and quasars can be measured using empirically 
derived \feii\ emission template spectra
\citep[WNW85;][]{Laetal97,Mcetal99,Dietal02a,Iwetal02}.
We fitted the \feii\ emission in our quasar spectra using scaled and broadened 
empirical \feii\ emission template spectra to derive relative \feii\ emission 
strength values.
For the ultraviolet wavelength range these templates had been carefully
extracted from {\it HST} observations of I\,Zw1 by \citet{VeWi01}. These
emission templates include contributions of \feii\ and \feiii .
In the fitting process the strength and the broadening of the Fe emission 
template are free parameters.

The use of different empirical \feii\ emission templates, in particular in the 
range of the broad \mgii $\lambda 2798$ emission line, introduces additional 
uncertainty to measured \feii /\mgii\ ratios.
For example,
the empirical Fe emission template we use in our analysis underestimates the 
Fe flux in the region. \citet{VeWi01} set the flux level beneath the broad 
\mgii $\lambda 2798$ emission line to zero in the Fe template extracted from 
IZw\,1. This approach is reasonable since detailed model calculations 
\citep{Veetal99,SiPr03} indicate that the Fe flux level in this wavelength 
range is of the order of only 15 to 20 \%\, compared to neighboring Fe 
emission. 
The treatment of the Fe emission beneath \mgii\ line has a major impact on the
\mgii\ flux. Since the broad component can contain $\sim 50 - 80$\,\%\ of the 
total line flux, by neglecting the presence of the broad profile component can 
easily cause an underestimate of the \mgii\ flux by a factor of $\sim 2$
\citep{VeWi01,Dietal02b}.

We explored the effects of non-zero \feii\ fluxes beneath the \mgii\ emission
line.
The gap in the Fe emission template, we used for the analysis \citep{VeWi01},
was
replaced by a low order polynomial ($2^{nd}$ order) which minimal flux level 
accounts to $\sim 15$\,\%\ of the immediately close by located Fe emission 
peaks. This choice is motivated by the photoionization calculations mentioned 
above \citep{Veetal99,SiPr03}. 
This additional flux increases the integrated \feii\ flux in the range 
$\lambda \lambda 2200 - 3090$\,\AA\ by less than 2\,\%.
Hence, the integrated \feii\ flux is nearly unchanged by filling the gap 
beneath the \mgii\ line with a more realistic estimate of the complex \feii\ 
emission.

To estimate the implications of the modified Fe emission template on the 
measured \mgii $\lambda 2798$ emission line strength we re-analyzed the 
\mgii\ line profile in each individual high-redshift and composite quasar 
spectra. The line profile was fit with a narrow and a broad Gaussian component.
The narrow component was nearly unchanged. The strength of the broad component 
was reduced by $\sim 6$\,\% on average. 
The cumulative effect of the slightly increased \feii\ emission and the reduced
strength of the \mgii\ emission flux amounts to $\sim 10$\,\%\ increase for the
measured \feii /\mgii\ ratios, nearly independent of the \mgii\ profile width.

\subsection{Strong Broad Emission Lines}
The broad emission line profile of \mgii $\lambda 2798$ was fit after
subtracting the power law continuum, the \feii\ emission, and the Balmer
continuum emission. 
Generally, the \mgii $\lambda 2798$ emission line profile could be 
reconstructed with two Gaussian components, 
a narrow component (FWHM(\mgii ) = $2270 \pm 570$ km\,s$^{-1}$)
and a blueshifted broad component ($\Delta v \simeq -1000$ km\,s$^{-1}$,
FWHM(\mgii ) = $6300 \pm 1300$ km\,s$^{-1}$).
The approach to reconstruct the \mgii\ emission line profile with two Gaussian 
components was only chosen to measure the line flux; each individual component 
has no physical meaning by itself.

\subsection{Internal Reddening}
In recent years growing evidence has appeared for the presence of large 
amounts of dust ($M_{dust} \gtsim 10^8$\,M$_\odot$) in the host galaxies 
of high-redshift quasars 
\citep[e.g.,][]{Guetal97,Guetal99,Caetal00,Ometal01}. It is 
assumed that the dust is distributed in a kiloparsec-scale warped disk 
\citep{Saetal89} which is illuminated by the central AGN. The observed 
dust emission spectra from 3 to 30 $\mu$m can be explained by such a 
model as shown by \citet{Anetal99} and \citet{Wietal00}.
However, since the presented spectra are typical quasar spectra with 
prominent broad emission lines, it is unlikely that the broad-line region
(BLR) is significantly blocked by dust \citep[e.g.,][]{NeLa93}.
Even if the radiation has to pass through an (external) dust screen the 
extinction of the \mgii\ and the UV \feii -emission (having about the 
same mean wavelength) will be comparable and the line ratio is not
expected to be significantly modified. 

\subsection{Multi-Component Fit of the Quasar Spectra}

We have analyzed the individual high-redshift quasar spectra, as well as the 
composite quasar spectra in a multi-component fit approach using emission 
templates as described above.
The spectra are transformed to their restframe using the corresponding 
redshift given in Table 2.
The quasar spectra were iteratively fitted first with a power law continuum, 
then a Balmer continuum emission template and an Fe emission template were
employed, and finally a two Gaussian profile fit was used to measure the broad
\mgii $\lambda 2798$ emission line.
The strength of the Fe emission templates was varied, as well as the width of 
the individual iron features.
The broadening of the empirical Fe emission template \citep{VeWi01}, was 
achieved by convolution with a Gaussian profile. The width of the iron emission
features of the best fit is consistent with the profile width of the 
\civ $\lambda 1549$ and \mgii $\lambda 2798$ emission lines to within 
$\sim 1000$ km\,s$^{-1}$.
We also varied the strength and the optical depth $\tau _\lambda$ of the Balmer
continuum emission ($0.1 \leq \tau _\lambda \leq 1.0$).
Minimal $\chi^2$ fits were calculated to determine the best fit. 
In order to estimate the accuracy of our fits, we also calculated fits for
settings of different parameters around the best fit values.
From the widths of the resulting $\chi ^2$ distributions we estimated mean 
errors for these parameters which were employed to estimate the uncertainties
of the flux measurements.
In spite of a signal-to-noise ratio of $S/N\simeq 10$ in the continuum of the
individual high-redshift quasar spectra, the UV \feii\ and \mgii\ emission 
strength can be measured to within $\sim 12 - 20$\,\%\ (\feii) and 
$\sim 10$\,\%\ (\mgii), respectively.
The errors of the \feii /\mgii\ ratio determined for the composite spectra
also take into account the uncertainties introduced by the individual spectra 
that contribute to the corresponding composite spectrum. We estimated this 
uncertainty from the rms-spectrum which is computed for each composite 
spectrum.

\section{Results}

In Figure 1 we present the quasar spectra of the $z = 4.5$ sample together with
the power law continuum fit, the appropriately scaled Fe and Balmer continuum
emission templates, and the fit of the \mgii $\lambda 2798$ emission line.
The achieved multi-component fit results benefit significantly from the large 
covered wavelength range to define a reliable continuum fit, in contrast to 
prior studies.
The residuum, subtracting the best fit from the quasar spectrum, is displayed 
in the bottom panel for each quasar. The remaining quite strong feature in the 
residuum spectra at $\lambda \simeq 3200$\,\AA\ is due to prominent iron 
emission blends, multiplets M6 and M7, that is not covered by the Fe template 
which we are using.
The ultraviolet \feii\ emission line flux was measured from the scaled Fe 
templates for a wavelength range of $\lambda \lambda 2200 - 3090$\,\AA. The 
integrated flux of \mgii $\lambda 2798$ was determined from the two component 
Gaussian fit.
The measured flux for both emission features are given in Table 2.

For SDSS\,1204$-$0021 only the H-band spectral range was observed.
Instead of applying an uncertain multi-component fit -- the continuum strength 
and the contribution of the Balmer continuum emission can not be reliably 
determined -- we compared the strength of the \mgii $\lambda 2798$ and the 
broad \feii\ emission feature at $\lambda \simeq 2500$\,\AA\ with the quasars 
at $z\simeq 4.5$. We calculated a mean $z\simeq 4.5$ quasar spectrum based on 
the other quasars of the $z=4.5$ sample.
In Figure 2 the spectrum of SDSS\,1204$-$0021 transformed to the restframe is
shown together with the mean $z\simeq 4.5$ quasar spectrum. This mean 
high-redshift quasar spectrum was scaled by multiplying with a constant to 
match the spectrum of SDSS\,1204$-$0021. The difference spectrum is displayed 
in the bottom panel of Figure 2. The overall spectral slopes of both these 
spectra are very similar.
Since the spectra were scaled to match the overall spectral shape a small
residuum is left of \mgii\ emission. The \mgii\ emission line profile is 
significantly broader in SDSS\,1204$-$0021 ($FWHM \simeq 7400$ km\,s$^{-1}$) 
than in the mean $z\simeq 4.5$ quasar spectrum ($FWHM \simeq 3000$ 
km\,s$^{-1}$) and it does not show a pronounced narrow component. 
Although the profiles of \mgii\ look quite different this is predominantly due 
to the missing narrow component in SDSS\,1204$-$0021, as can be seen in the 
difference spectrum (Fig.\,2, bottom panel). The residual flux
in the difference spectrum amounts to $\sim 5$\,\%\ of the total \mgii\ flux
in both quasar spectra. Although the \feii /\mgii\ ratio can not be measured
in a multi-component fit, the comparison with the mean $z\simeq 4.5$ quasar
spectrum indicates that the ratio should be very similar to the mean value of
the individual quasars at $z>4$.

\subsection{\feii/\mgii\ Ratio as a Function of Redshift}

First, we compare the \feii/\mgii\ ratios for the $z = 3.4$ and $z = 4.5$ 
quasar sample we observed for individual quasars. The measured \feii/\mgii\
ratios which are given in Table 2, are displayed as function of redshift in 
Figure 3. 
The results for the quasars of the $z=3.4$ sample \citep{Dietal02a} are 
supplemented with Q\,2050$-$359 ($z=3.51$) which we observed at CTIO. 
There is obviously no strong indication for an evolution of the \feii /\mgii\ 
ratio from $z\simeq 3.2$ up to $z\simeq 4.8$. Including the result of the 
comparison of the SDSS\,1204$-$0021 spectrum with the mean  $z = 4.5$ quasar 
spectrum, the \feii/\mgii\ ratio shows no variation up to $z\simeq 5$.
Although our observation of PC\,1158$+$4635 indicates a lower \feii/\mgii\
ratio, SDSS\,1204$-$0021 and recent NICMOS/HST observations of three
$z\simeq 6$ quasars show no declining \feii/\mgii\ ratio 
\citep{Freu03}. 

To investigate the evolution of the \feii /\mgii\ ratio over a wide range in
cosmic time, particularly from the local universe to $z\simeq 5$, we measured
this emission line ratio in the quasar spectra of the redshift sample.
The criterion of nearly constant and also comparable luminosity of the
composite spectra of this sample (Table 2) compared to the $z=3.4$ and $z=4.5$
quasar samples minimizes luminosity effects on the emission line strength like 
the Baldwin effect \citep[e.g.,][]{OsSh99,Cretal02,Dietal02b,Shetal03}.
The \mgii $\lambda 2798$ and UV \feii\ line fluxes we measure for the quasars 
of the redshift sample are given in Table 2.
In Figure 3 these \feii/\mgii\ ratios are plotted together with the results for
the individual high-redshift quasars.
The analysis of the redshift quasar spectra provides further evidence that the 
ratio of iron to $\alpha$-elements expressed by \feii/\mgii\ lacks an evolution
for luminous quasars from the local universe ($z\simeq 0$) up to $z\simeq 5$, 
i.e., an cosmic age of $\sim 1$\,Gyr (Fig.\,3).

\subsection{\feii/\mgii\ Ratio as a Function of Luminosity}

To investigate the dependence of the \feii/\mgii\ ratio on luminosity we 
analyzed the composite quasar spectra of the luminosity sample which cover 
nearly $\sim 6$ orders of magnitude in luminosity \citep{Dietal02b}.
In Figure 4 the \feii/\mgii\ ratios measured for these composite spectra 
(luminosity sample) are displayed as a function of continuum luminosity 
$\lambda L_\lambda (1450)$. In addition, the \feii/\mgii\ ratios of the 
individual high-redshift quasars of the $z=3.4$ and $z=4.5$ samples are shown.
Within the errors there is a weak trend for a higher \feii/\mgii\ ratio towards
higher luminosities (Fig.\,4). It appears that the ratio is nearly constant for
$log \lambda\,L_\lambda (1450{\rm \AA}) \la 45$ erg\,s$^{-1}$ with 
\feii/\mgii $=2.1\pm0.3$ and amounts to \feii/\mgii $=4.0\pm0.8$
at higher luminosities.
The composite quasar spectrum at $log \lambda\,L_\lambda = 43.51$ erg\,s$^{-1}$
consists of only 3 individual spectra from which two cover the UV 
\feii\ -- \mgii\ wavelength range. One of these two quasars is WPVS007, a 
well-known narrow-line Seyfert\,1 (NLS1) galaxy. The {\it HST} spectrum that we
retrieved from the {\it HST} data archive, shows exceptionally strong \feii\ 
emission \citep{Gretal95} which is a common characteristic of NLS1 galaxies 
\citep[e.g.,][]{Mathur00,Pogge00}. Thus WPVS007 dominates the composite 
spectrum at $log \lambda\,L_\lambda = 43.51$ erg\,s$^{-1}$ and results to the
high value of \feii /\mgii\ = $5.6\pm1.2$.

\subsection{Comparison with other Studies}

During the last few years there have been a few similar studies of the UV 
\feii/\mgii\ ratio in quasars at high redshifts \citep[e.g.,][]{Kaetal96, 
Taetal97,Muetal99,Thetal99,Iwetal02,Freu03}, as well as for lower redshift 
samples \citep{Wietal80,Gran81,WNW85}.
However, to compare those results with the \feii/\mgii\ ratios we measure 
here ($z=3.4$, $z=4.5$, and redshift and luminosity quasar samples), it is 
necessary to adjust the published ratios to reflect the same spectral range we
used to measure the \feii\ emission (see below).
The comparison is further complicated by the different \feii\ emission
templates which were used to measure the \feii\ flux, as well as taking into
account the contribution of the Balmer continuum.
Wills et al.\,(1985) applied an \feii\ emission template based on their 
photoionization model calculations \citep{NeWi83}, while 
\citet{Thetal99} and \citet{Iwetal02} used the LBQS mean quasar 
spectrum \citep{Fraetal91} to extract an empirical \feii\ emission 
template.
Another more serious uncertainty is introduced by the generally restricted 
wavelength coverage of prior studies that made a proper definition of the 
underlying continuum strength very difficult, in particular, in the restframe 
UV wavelength range of high-redshift quasars.
The quasar spectra Wills et al.\,(1985) analyzed covered at least the spectral
range from \ciii]$\lambda 1909$ up to H$\beta\,\lambda 4861$.
The high redshift quasars ($z=3.4$ and $4.5$ samples) in the \citet{Thetal99} 
study cover $\sim 600$\,\AA\ around the \mgii $\lambda 2798$ emission line
and, hence, does not cover the entire \feii\ bump. The
individual spectra of the Thompson et al. sample have been re-analyzed by 
\citet{Iwetal02},
who also observed the \mgii\ -- UV \feii\ wavelength range for 13 quasars with 
$z\ga 4.4$, across a wavelength range of $\sim 1200$\,\AA . \citet{Iwetal02}
also presented measurements of \feii/\mgii\ for the early data release of the 
SDSS QSO survey, as well as for quasar spectra retrieved from the {\it HST} 
and {\it IUE} data archives for a local quasar population.
Recently, \citet{Freu03} presented results of measured \feii/\mgii\ ratios
for three quasars at $z\simeq 6$. Although the uncertainties are quite large
the \feii/\mgii\ ratios are comparable to the results of the $z\ga 4$ quasars
both here and by \citet{Iwetal02}.

To reduce the most obvious source of differences for the individual published 
\feii/\mgii\ ratios we adjusted the \feii/\mgii\ emission line ratios of 
prior studies to the integration window that we used to measure the \feii\ 
emission line flux. Therefore, we compared the Fe emission flux we measured for
the integration range we used ($\lambda \lambda 2200 - 3090$\,\AA ) with the 
those resulting by applying different integration windows.
\citep[WNW85;][]{Thetal99,Iwetal02}.

In Figure 5 we present our measurements of the \feii/\mgii\ ratio for the 
$z=3.4$, $z=4.5$, and redshift samples in comparison with the rescaled 
results presented by \citet{WNW85}, \citet{Thetal99}, \citet{Iwetal02}, and
\citet{Freu03}.
At higher redshifts ($z\ga 3$) the intercalibrated \feii /\mgii\ ratios are 
quite consistent although the scatter of the individual measurements given by 
\citet{Iwetal02} is large. 
In particular, the \feii/\mgii\ ratios for the high-redshift ($z\ga 4$) 
composite spectra of the SDSS sample which are also those with the highest 
luminosity of this sample \citep{Iwetal02} agree quite well with the results
of our composite spectra (\feii /\mgii\,$\simeq 4.4 ~{\rm to}~ 5.7$ compared to
$5.4\pm1.3$ here, Table 2).
The \feii/\mgii\ measurements for three quasars at the highest redshift 
indicate that comparable high \feii/\mgii\ ratios can be found up to redshifts 
of $z\simeq 6$ \citep{Freu03}.
However, the scatter of the \feii/\mgii\ ratios at $z\ga 4$ given in these
studies is quite large. We presume that a significant fraction of the large 
range of these measured \feii /\mgii\ ratios is caused by the restricted
wavelength coverage of the analyzed quasar spectra. Although the mathematically
best solution was determined for the multi-component fit, the continuum 
strength appears in some cases not well defined, yielding quite
uncertain \feii\ flux measurements. 
Some scatter in the \feii/\mgii\ ratio given in recent studies may also be 
introduced by using emission lines that are not completely covered by the 
available spectra or measured at all, instead has been estimated from the
\ciii]$\lambda 1909$ emission blend \citep{Freu03}.

In general, at lower redshifts ($z\la 2$) the \feii/\mgii\ ratios derived from 
the SDSS quasar spectra ($2.6\pm 0.3$) are in good agreement with the ratios we
measured ($2.1\pm0.3$) from the composite spectra of $z\la 1$ quasars with 
lower luminosities (see Fig.\,4, luminosity sample, 
$log \lambda\,L_\lambda (1450{\rm \AA}) \la 45$ erg\,s$^{-1}$).
The sample of quasars studied by Wills et al.\,(1985) with $z\la 0.8$ infer
\feii/\mgii\ ratios which also are quite consistent with the ratios we derived.
However, the intercalibrated \feii/\mgii\ ratios based on the SDSS quasars are 
significantly lower than the ratios we determined from composite spectra of 
nearly constant luminosity for redshifts less than $z\simeq 2$ (Fig.\,5).
A possible explanation may be given by the lower luminosity of these SDSS
quasars compared to our composite quasar spectra in this redshift range
with $log \lambda\,L_\lambda (1450{\rm \AA}) \simeq 46.7$ erg\,s$^{-1}$ for
the following reasons.
Although the \mgii $\lambda 2798$ and \feii\ emission may both originate in the
partially ionized zone of the line emitting region, both lines depend in
different ways on the conditions of this region, e.g., radiative transfer,
destruction by Balmer continuum absorption, and conversion to optical \feii\ 
emission \citep[e.g.,][]{WNW85,Joly87,Veetal99}.
Luminosity dependent variations in the SED of the AGN continuum, 
in particular of the ionizing continuum, are known to exist based on 
observational and theoretical studies \citep[e.g.,][]{NeLaGo92,Laoetal95,
Wand99,Dietal02b}. Hence, quasars of different luminosity may have a different 
continuum SED which may result in slightly different conditions of the \feii\ 
and \mgii $\lambda 2798$ emitting region ($z\la 2$). Taking the sensitive 
dependence of \feii\ and \mgii $\lambda 2798$ on the conditions of the gas into
account this may result in a wide range of possible \feii /\mgii\ ratios as 
observed at low redshifts ($z\la 0.2$).
The large scatter of the \feii /\mgii\ ratio also may emphasize that it is
important to compare this line ratio for quasars of similar luminosity.

\section{Discussion}

The \feii /\mgii\ emission line ratios which we find for our high-redshift 
quasar sample, as well as for the composite quasar spectra are consistent with 
results presented by Wills et al.\,(1985), \citet{Thetal99}, \citet{Muetal99}, 
\citet{Dietal02a}, \citet{Iwetal02}, and \citet{Freu03}.
In particular, the \feii/\mgii\ ratio at high redshifts shows similar strength
compared to those measured in low redshift quasars ($z\simeq 0.2$) of 
comparable luminosity
(WNW85). Assuming that the physical conditions in the BELR of quasars and that
the spectral energy distribution of the ionizing continuum are independent of
redshift the observed lack of evolution of the \feii/\mgii\ emission ratio
suggests that the observed relative abundance of iron, i.e. Fe/Mg, in the local
universe is already achieved at redshifts $z\ga 4$.
This result indicates that an episode of intense star formation began before 
the observed quasar activity.
Hence, the nearly constant \feii/\mgii\ ratio over cosmic time has important
implications on the star formation history, particularly at high redshifts
since the age of the universe is less than $\sim 1.3$\,Gyr (for redshifts 
$z\ga 4.5$).

With this result, we can estimate the beginning of the first star
formation in luminous quasar host galaxies, i.e., probably young massive 
elliptical galaxies from the evolution time scale of type Ia SNe and 
the resulting delay in the Fe enrichment.
Evolutionary models of the Fe enrichment in a galaxy following the initial 
starburst has been calculated e.g. by \citet{Yoetal98}. Although, as pointed 
out by Yoshii et al., evidence that the initial mass function may have been 
different for the first stars makes such model calculations quantitatively
somewhat uncertain, it is clear from these computations that the relative Fe 
abundance is rather low initially and starts to grow steeply about $\sim 1$ Gyr
after the beginning of the star formation, reaching a maximum (at about 3 Gyrs
in the Yoshii et al. model), before declining to the local value 
\citep[see also ][]{MaPa93,HaFe93}. 
The same temporal evolution of the Fe/Mg ratio is found for the giant 
elliptical galaxy model (M4a) presented by \citet{HaFe93}. Their model
predicts a strong increase of Fe/H at $\sim 1$\,Gyr after the beginning 
of the star formation, with the most rapid rise of the Mg/Fe ratio during 
the 1 - 1.38 Gyr period.

However, \citet{Matt94}, \citet{FrTe98}, and \citet{MaRe01} pointed out that 
the common assumption $\tau_{delay} \simeq 1$ Gyr as a typical time scale for 
the enrichment delay of Fe, released by SNe type Ia SNe explosions, is only a 
rough estimate which based on the conditions of the solar neighborhood.
The evolutionary models presented by \citet{FrTe98} and \citet{MaRe01} suggest 
that this delay depends crucially on environmental conditions of the star 
formation process and that it can be significantly shorter.
They showed that for elliptical galaxies an intense early star formation 
episode with enhanced star formation rate and short star formation times scale
results in a high SNIa rate which reaches a maximum already $\sim 0.3 - 0.8$ 
Gyr after the beginning of the star formation, in contrast to 
$\sim 4\,{\rm to}\,5$ Gyr as for the solar neighborhood.
Various theoretical studies have explored possible scenarios that SNe type Ia 
can, under certain conditions, also be produced by massive stars with a much 
shorter evolutionary time scale than that assumed in the standard scenarios
\citep[e.g.,][]{Yoetal98}, some SNe type Ia progenitors could have much shorter
life times \citep[e.g.,][]{IbTu84,SmWy92}. 

There is growing evidence that luminous quasars are residing in massive host 
galaxies \citep[e.g.,][]{Kuetal01,Noetal01,Duetal03}, which at high redshift
($z\ga 3$) are probably young ellipticals or massive spheroidal systems in the
process of forming \citep[e.g.,][]{vanBretal98,Papaetal00,deBretal02}.
This would support the assumption of significantly shorter evolutionary time 
scales for the major iron enrichment in massive spheroidal galaxies than that
generally cited, $\tau _{evol} \simeq 1$ Gyr.
As noted above, the significant release of iron by type Ia SNe to the 
interstellar medium can be as short as $\tau _{evol} \la 0.3 - 0.8$ Gyr. 
In addition, a more rapid evolution of progenitor stars to produce and release 
iron to the interstellar environment than previously assumed may be also 
possible. 
Recent models of galaxy formation, particularly in the early universe suggest
that a rapid assembly of (at least some) massive spheroidal systems was 
accompanied by intense star formation, and subsequent quasars activity 
\citep[e.g.,][]{KaHa00,Gretal01,Roetal02}. The 
indication of an early star formation epoch beginning at $z_f \simeq 6 - 9$
is also consistent with cosmic structure formation models \citep{GnOs97} and 
with recent estimates of the epoch of reionization.

The observed lack of an evolving \feii/\mgii\ ratio indicates that the star 
formation responsible for the chemical enrichment of the gas close to the 
quasar has started at least $\tau _{evol}$ earlier.
Therefore, the first major star formation epoch in high-redshift quasars
($z\ga 4.5$) started at $z_f \simeq 6 - 9$. 
It is interesting to note that this estimate is comparable to the epoch of 
re-ionization of the universe \citep{(GnOs97,HaLo98,Becketal01,Fanetal02}. 
The indication of an early intense star
formation epoch is further supported by recent results of the Wilkinson 
Microwave Anisotropy Probe (WMAP) \citep{Beetal03}. \citet{Cen03} suggests 
that the first stars were formed as early as $z\ga 15$ and that they re-ionize 
the universe for the first time, while a second re-ionization episode of the 
universe occurred at $z\simeq 6$. 

\section{Conclusion}
We have observed 6 luminous high-redshift quasars ($\ga 4.4$) in the near
infrared wavelength range to measure the \feii/\mgii\ emission line ratio
and thereby estimate the beginning of the first intense star formation. 
To explore the evolution of the \feii/\mgii\ ratio over a wide redshift range,
i.e. cosmic time, we measured this ratio for composite quasar spectra which 
cover a redshift range of $0 \la z\la 5$ with nearly constant continuum 
luminosity (redshift sample), as well as for composite quasar spectra which 
span nearly $\sim 6$ orders of magnitude in luminosity (luminosity sample).
We find that the \feii/\mgii\ ratio shows no evolution for cosmic times 
corresponding $3.2 \la z \la 5.0$ compared to measurements for quasars at 
$z=\simeq 3.4$ \citep[see e.g.,][]{Dietal02a}.
The \feii/\mgii\ ratio also shows no evolution from the local universe to very
high redshifts ($z\simeq 5$), corresponding to an age of the universe of 
$\sim 1$ Gyr. 
For increasing continuum luminosities, however, we detect a weak trend for 
higher \feii/\mgii\ ratio at higher luminosities. It appears that at lower 
luminosities ($log\,\lambda L_\lambda (1450)\la 10^{45}$\,erg\,s$^{-1}$) 
\feii/\mgii\,$\simeq 2$ while at higher luminosities 
($log\,\lambda L_\lambda (1450)\ga 10^{46}$\,erg\,s$^{-1}$)
\feii/\mgii\,$\simeq 4 - 5$.

Assuming that luminous quasars reside in massive galaxies, probably elliptical
galaxies or in forming massive spheroidal systems in the early universe 
\citep[e.g.,][]{KaHa00,Kuetal01,Noetal01,Roetal02,Duetal03}
the lack of evolution in the observed \feii/\mgii\ ratios across the redshift
range $0 \la z \la 5$ suggests that there is also no evolution in Fe/Mg
in the quasar nuclear region.
The actual value of Fe/Mg in quasar BELRs remains uncertain. Further modeling
of the \feii\ and \mgii\ line emission is required. However, there is some
evidence that Fe/Mg is above solar in quasars at low to moderate redshifts
(Wills et al.\,1985). If that result is confirmed, the lack of evolution in 
\feii/\mgii\ out
to $z\simeq 5$ reported here would indicate that the first major star formation
in the environment of high-redshift quasars started at $z_f \simeq 6 - 9$,
corresponding to an age of the universe of $\sim 0.5$ Gyr.
It is interesting to note that this epoch is comparable to the proposed 
time of the re-ionization of the universe \citep{HaLo98,Cen03}.

\acknowledgments{ 
This work was supported by NASA through their Long Term Space Astrophysics 
program (NAG5-3234), NSF AST99-84040, and an archival research grant from the 
Space Telescope Science Institute (AR-07988.01-96A).
MV acknowledges financial support from the Columbus Fellowship and support for 
Proposal number AR-09549, provided by NASA through a grant from the Space 
Telescope Science Institute, which is operated by the Association of 
Universities for Research in Astronomy, Incorporated, under NASA contract 
NAS5-26555.)

\newpage
\figcaption[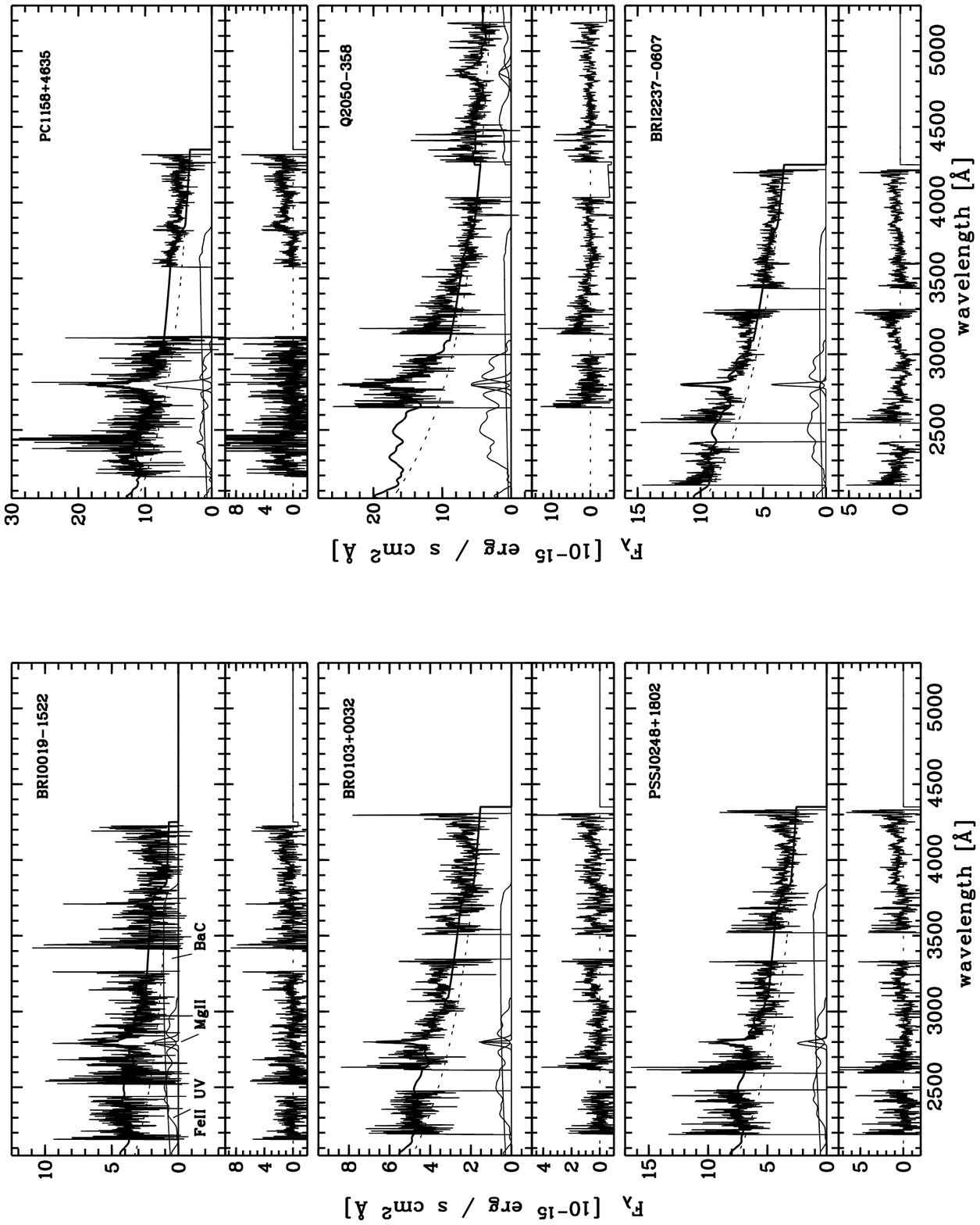]{The restframe quasar spectra together with 
                    results of the multi-component analysis are shown for 
                    BRI\,0019$-$1522, BR\,0103$+$0032, PSSJ\,0248$+$1802, 
                    PC\,1158$+$4635, Q\,2050$-$358, and BRI\,2237$-$0607.
                    In the top panel the quasar spectrum is shown together 
                    with the power law continuum fit (dotted line), the scaled
                    and broadened Fe-emission template, the scaled Balmer 
                    continuum emission, and the Gaussian components to fit
                    the \mgii $\lambda 2798$ emission line profiles. The 
                    resulting fit is over plotted as a thick solid line.
                    In the bottom panel the quasar spectrum is shown after 
                    subtraction of these components.}

\figcaption[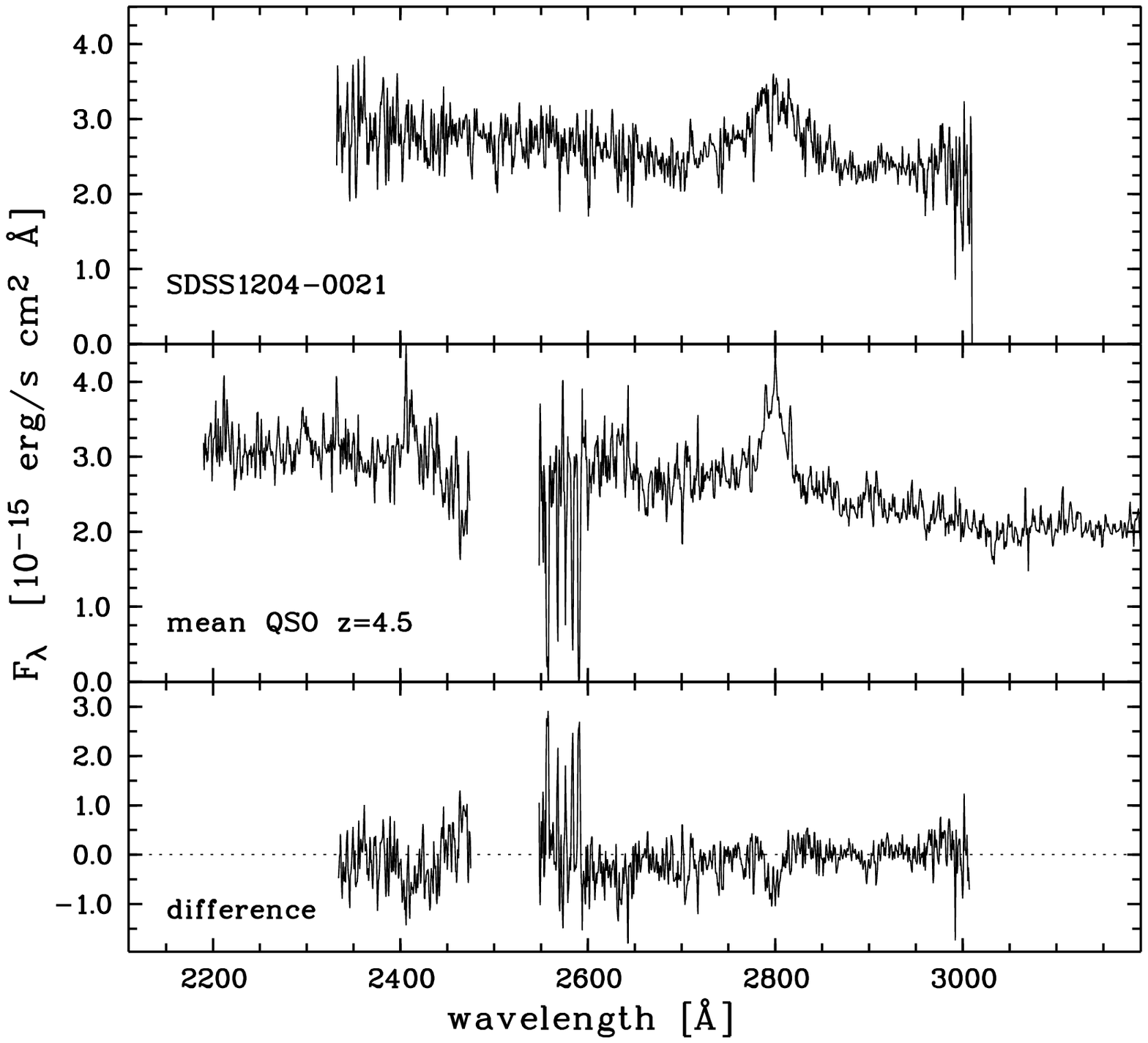]{The spectrum of SDSS\,1204$-$0021, observed in the
                    H-band 
                    spectral region, transformed to the restframe (top panel).
                    In the middle panel the mean $z\simeq 4.5$ quasar spectrum
                    is shown, scaled to match the overall spectral shape of
                    SDSS\,1204$-$0021 and the resulting difference spectrum is 
                    plotted in the bottom panel.}

\figcaption[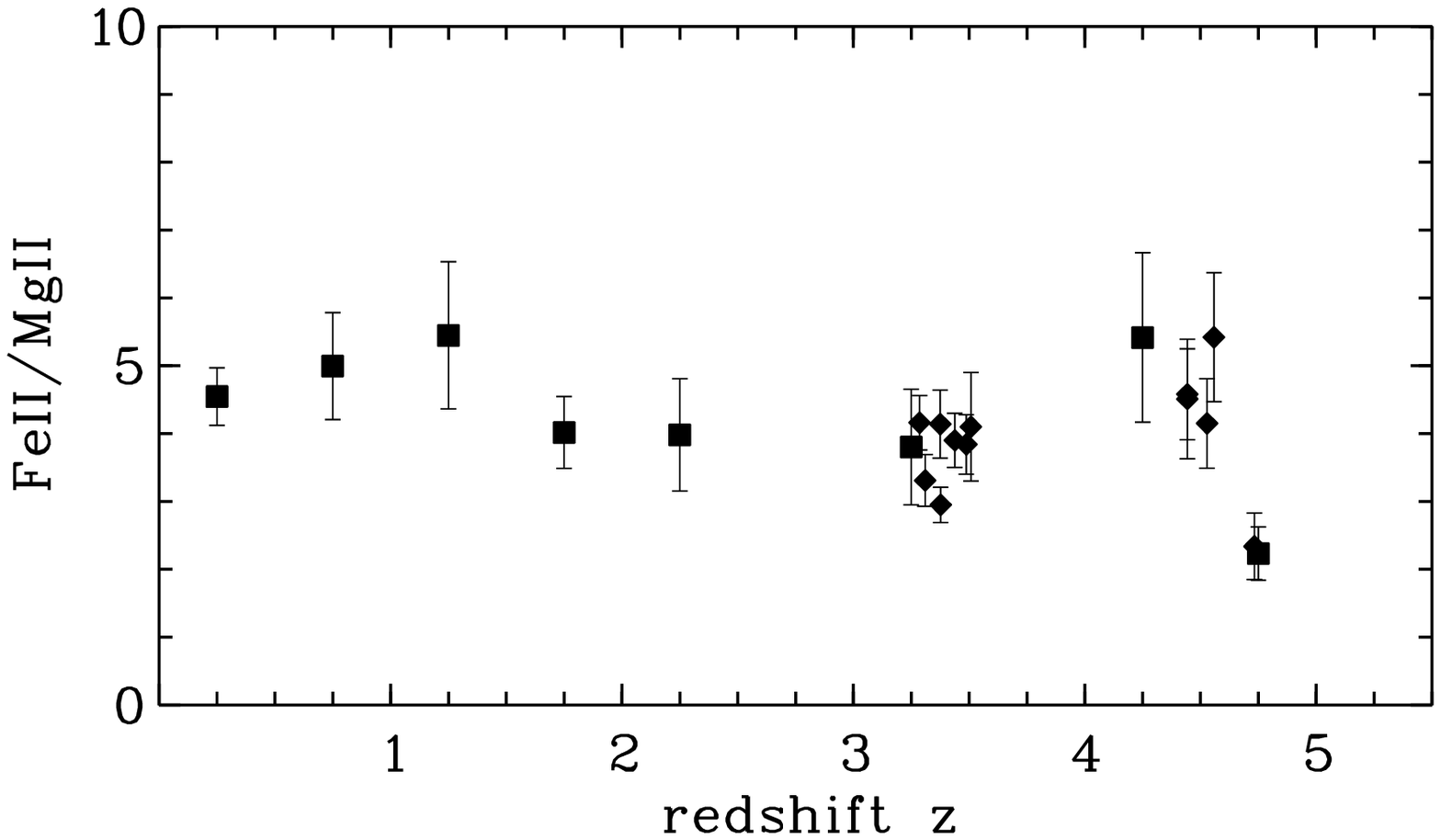]{The \feii/\mgii\ line ratio of the $z\simeq 4.5$ quasars 
                     in comparison with quasars at $z\simeq 3.4$ which have 
                     been observed previously \citep{Dietal02a} are
                     plotted as filled diamonds as a function of redshift.
                     The \feii/\mgii\ ratio of composite spectra (filled boxes,
                     redshift sample) and comparable luminosity 
                     like the individual high-redshift quasars are displayed to
                     illustrate the lack of evolution of the \feii /\mgii\ 
                     line ratio.}

\figcaption[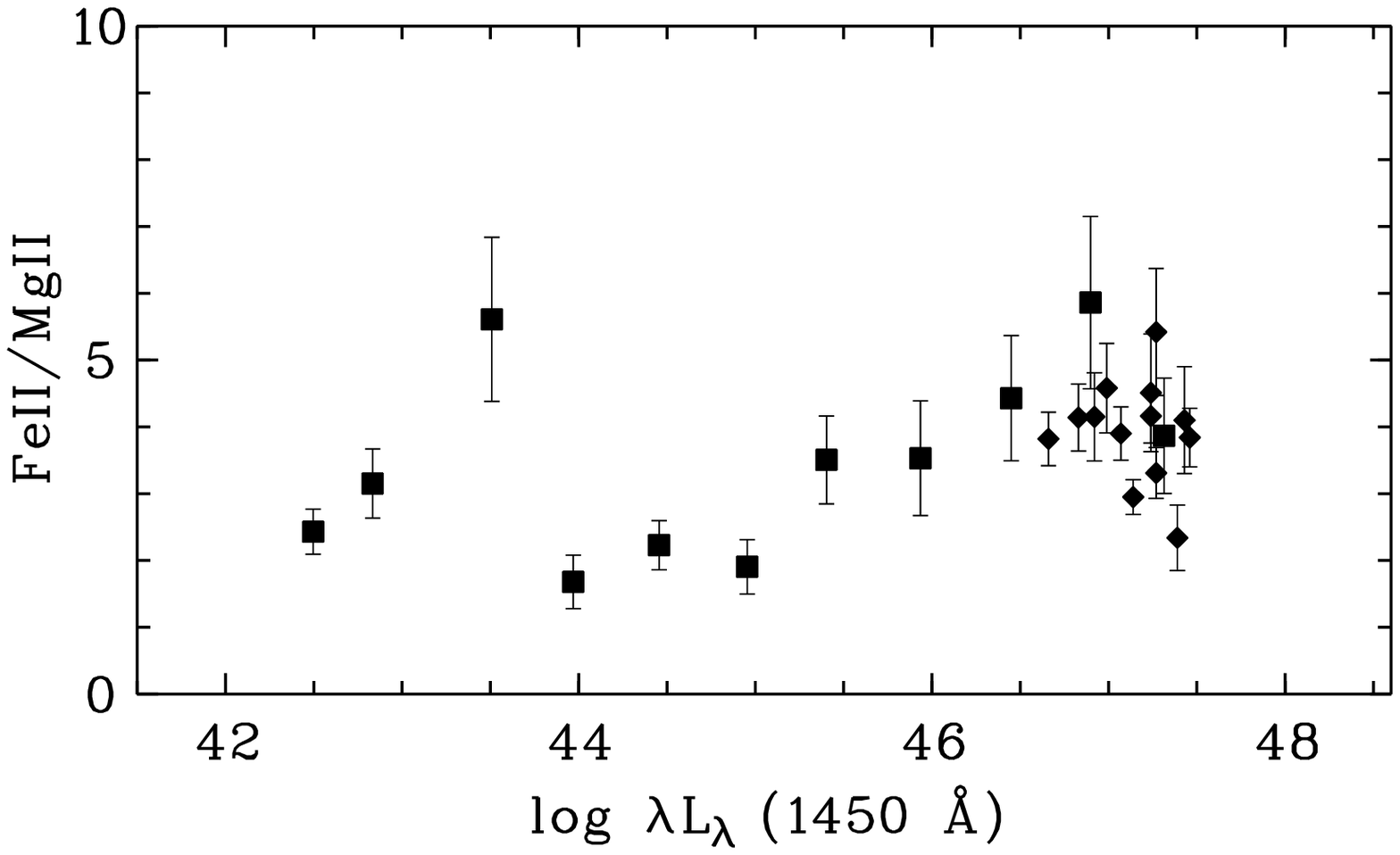]{The same as Figure 3 but the \feii/\mgii\ ratios
                    of the individual high-redshift quasars (filled diamonds) 
                    and 
                    composite spectra with increasing luminosity (filled boxes,
                    luminosity sample)
                    are plotted as a function of intrinsic continuum 
                    luminosity, $\lambda L_\lambda(1450{\rm \AA})$.}

\figcaption[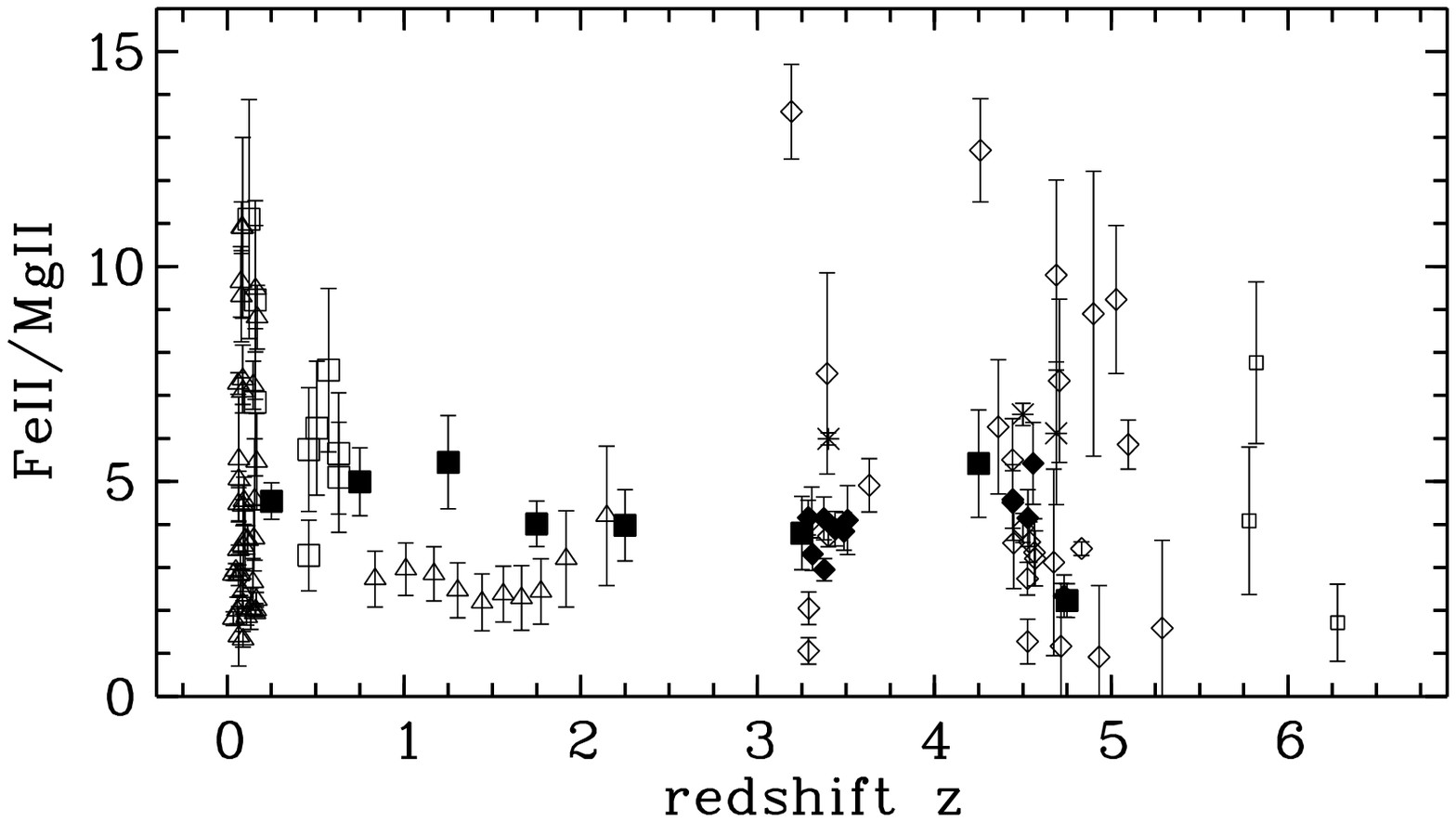]{The same as Figure 3 but the \feii/\mgii\ ratios
                    of the $z=3.4$ and $z=4.5$ high-redshift quasars 
                    (filled diamonds $\blacklozenge$) and redshift sample
                    quasar spectra (filled boxes $\blacksquare$) of our study 
                    are compared to the results of \citet{WNW85} $\sq$,
                    \citet{Thetal99} $\ast$, \citet{Iwetal02}
                    ($\diamond$, $\Delta$), and \citet{Freu03} 
                    (small $\sq$).}

\clearpage

%
\begin{figure}
\plotone{f1.eps}
\end{figure}
\begin{figure}
\plotone{f2.eps}
\end{figure}
\begin{figure}
\plotone{f3.eps}
\end{figure}
\begin{figure}
\plotone{f4.eps}
\end{figure}
\begin{figure}
\plotone{f5.eps}
\end{figure}
%


\begin{deluxetable}{lccccc}
\tablewidth{0pt}
\tablecaption{The high redshift quasars sample}
\tablehead{
\colhead{quasar} &
\colhead{site} &
\colhead{date} &
\multicolumn{3}{c}{exposure time}\\ 
\colhead{ } &
\colhead{ } &
\colhead{ } &
\colhead{J} &
\colhead{H} &
\colhead{K} \\
\colhead{ } &
\colhead{ } &
\colhead{ } &
\colhead{ } &
\colhead{[sec]} &
\colhead{ } 
}
\startdata
BRI 0019$-$1522 &CTIO&16-Sep-00      &2100&2100&2100\\
BRI 0103$+$0032 &CTIO&16,17-Sep-00   &8700&8700&8700\\
PSS J0248$+$1802&CTIO&17-Sep-00      &7200&7200&7200\\
PC 1158$+$4635  &Keck&20-May-00      &2400&2800&1920\\
SDSS 1204$-$0021&VLT &11-Apr-01      &\nodata&1800&\nodata\\
                &    &14-Jun-01      &\nodata&1800&\nodata\\
Q 2050$-$359    &CTIO&17-Sep-00      &5700&5700&5700\\ 
BRI 2237$-$0607 &CTIO&15,16,17-Sep-00&15800&15800&15800\\
\enddata 
\end{deluxetable}

\clearpage


\begin{deluxetable}{lcccc}
\tablewidth{0pt}
\tablecaption{\mgii$\lambda 2798$ and \feii\ emission line flux measurements of
              the $z=3.4$, $z=4.5$ quasar samples and of the redshift and 
              luminosity samples.}
\tablehead{
\colhead{quasar} &
\colhead{$z$} &
\colhead{$log \lambda\,L_\lambda (1450{\rm \AA})$} &
\colhead{F(\mgii)\tablenotemark{a}} &
\colhead{UV \feii/\mgii} \\
\colhead{ } &
\colhead{ } &
\colhead{[erg\,s$^{-1}$]} &
\colhead{ } &
\colhead{ } 
}
\startdata
\multicolumn{5}{c}{The z=3.4 Sample}\\
\hline
Q\,0103$-$260    &3.375&46.83&$ 73.6\pm3.5$&$4.14\pm0.50$\\
Q\,0105$-$2634   &3.488&47.46&$267.4\pm11$&$3.84\pm0.44$\\
Q\,0256$-$0000   &3.377&47.14&$200.7\pm 7$&$2.95\pm0.26$\\
Q\,0302$-$0019   &3.286&47.24&$341.5\pm10$&$4.16\pm0.40$\\
Q\,2050$-$359    &3.508&47.43&$566.5\pm70$&$4.10\pm0.80$\\
Q\,2227$-$3928   &3.438&47.07&$156.0\pm 5$&$3.90\pm0.40$\\
Q\,2348$-$4025   &3.310&47.27&$155.6\pm10$&$3.31\pm0.38$\\
\hline
\multicolumn{5}{c}{The z=4.5 Sample}\\
\hline
BRI\,0019$-$1522 &4.528&46.92&$135.9\pm15$&$4.15\pm0.66$\\
BR\,0103$+$0032  &4.443&46.99&$ 85.0\pm 8$&$4.58\pm0.67$\\
PSSJ\,0248$+$1802&4.443&47.24&$134.3\pm13$&$4.51\pm0.88$\\
PC\,1158$+$4635  &4.733&47.39&$394.8\pm40$&$2.34\pm0.49$\\
BRI\,2237$-$0607 &4.558&47.27&$138.5\pm14$&$5.42\pm0.95$\\
\hline
\multicolumn{5}{c}{The Luminosity Sample}\\
\hline
comp.L40l      &0.003&42.50&$1003\pm75$&$2.43\pm0.34$\\
comp.L40h      &0.034&43.51&$ 263\pm10$&$5.61\pm1.23$\\
comp.L41l      &0.051&43.97&$ 273\pm13$&$1.68\pm0.40$\\
comp.L41h      &0.088&44.46&$ 258\pm 9$&$2.23\pm0.37$\\
comp.L42l      &0.267&44.95&$ 105\pm 3$&$1.91\pm0.41$\\
comp.L42h      &0.721&45.40&$  38\pm 2$&$3.51\pm0.66$\\
comp.L43l      &1.349&45.94&$  60\pm 2$&$3.53\pm0.86$\\
comp.L43h      &2.069&46.45&$  81\pm 3$&$4.43\pm0.94$\\
comp.L44l      &2.780&46.90&$ 131\pm 4$&$5.86\pm1.29$\\
comp.L44h      &2.909&47.32&$ 312\pm21$&$3.87\pm0.86$\\
\hline
\multicolumn{5}{c}{The Redshift Sample}\\
\hline
comp.red025    &0.24 &46.29&$1278\pm36$&$4.54\pm0.43$\\
comp.red075    &0.80 &46.43&$ 249\pm 6$&$4.99\pm0.79$\\
comp.red125    &1.28 &46.56&$ 203\pm 6$&$5.45\pm1.08$\\
comp.red175    &1.87 &46.66&$ 154\pm 3$&$4.02\pm0.53$\\
comp.red225    &2.14 &46.70&$ 122\pm 7$&$3.98\pm0.83$\\
comp.red275    &2.79 &46.73&\nodata    &\nodata      \\
comp.red325    &3.17 &46.74&$ 72\pm 5$ &$3.80\pm0.85$\\
comp.red375    &3.84 &46.82&\nodata    &\nodata      \\
comp.red425    &4.26 &46.76&$ 64\pm 5$ &$5.42\pm1.25$\\
comp.red475    &4.79 &46.65&$ 77\pm 8$ &$2.23\pm0.39$\\
\enddata
\tablenotetext{a}{[$10^{-15}$ erg\,s$^{-1}$\,cm$^{-2}$]}
\end{deluxetable}


\begin{thebibliography}{}
\bibitem[Andreani et al.(1999)]{Anetal99}
Andreani, P., Franceschini, A., \& Granato, G. 1999, \mnras, 306, 161
\bibitem[Arnaud et al.(1989)]{Arnetal89}
Arnaud, K.A., Gilmore, G., \& Cameron, A.C. 1989, \mnras, 237, 495
\bibitem[Bautista \& Pradhan(1998)]{BaPr98}
Bautista, M.A. \& Pradhan, A.K. 1998, \apj , 492, 650
\bibitem[Becker at al.(2001)]{Becketal01}
Becker, R.H., et al. 2001, \aj, 122, 2850
\bibitem[Bennett et al.(2003)]{Beetal03}
Bennett, C.L., et al. 2003, \apj, 583, 1
\bibitem[Carilli et al.(2000)]{Caetal00}
Carilli, C.L., Bertoldi, F., Menten, K.M., et al. 2000, \apj, 533, L13
\bibitem[Carilli et al.(2002)]{Caetal02}
Carilli, C.L., Kohno, K., Kawabe, R., et al. 2002, \aj, 123, 1838
\bibitem[Carroll et al.(1992)]{CPT92}
Carroll, S.M., Press, W.H., \& Turner, E.L. 1992, \araa , 30, 499
\bibitem[Cen(2003)]{Cen03}
Cen, R. 2003, \apj, in press, (astro-ph/0303236)
\bibitem[Cen \& Ostriker(1999)]{CeOs99}
Cen, R. \& Ostriker, J.P. 1999, \apj, 519, L109
\bibitem[Collin \& Joly(2000)]{CoJo00}
Collin, S. \& Joly, M. 2000, New Astron.Rev., 44, 531
\bibitem[Connolly et al.(1997)]{Conetal97}
Connolly, A.J., Szalay, A.S., Dickinson, M., Subbarao, M., \& Brunner,
R.J. 1997, \apj, 486, L11
\bibitem[Corbin \& Boroson(1996)]{CoBo96}
Corbin, M.R. \& Boroson, T.A. 1996, \apjs, 107, 69
\bibitem[Croom et al.(2002)]{Cretal02}
Croom, S.M., et al. 2002, \mnras, 337, 275
\bibitem[De\,Breuck et al.(2002)]{deBretal02}
De\,Breuck, C., van Breugel, W., Stanford, S.A., R\"{o}ttgering, H., Miley, G.,
 \& Stern, D. 2002, \aj, 123, 637
\bibitem[Dietrich et al.(1999)]{Dietal99}
Dietrich, M., Appenzeller, I., Wagner, S.J., et al. 1999, \aap , 352, L1
\bibitem[Dietrich \& Wilhelm-Erkens(2000)]{DiWi00}
Dietrich, M. \& Wilhelm-Erkens, U. 2000, \aap , 354, 17
\bibitem[Dietrich et al.(2002a)]{Dietal02a}
Dietrich, M., Appenzeller, I., Vestergaard, M. \& Wagner, S.J. 2002a, \apj, 
564, 581
\bibitem[Dietrich et al.(2002b)]{Dietal02b}
Dietrich, M., et al. 2002b, \apj, 581, 912 
\bibitem[Dietrich et al.(2003a)]{Dietal03a}
Dietrich, M., et al. 2003a, \aap, 398, 891
\bibitem[Dietrich et al.(2003b)]{Dietal03b}
Dietrich, M., et al. 2003b, \apj, 589, 722
\bibitem[Dunlop et al.(2003)]{Duetal03}
Dunlop, J.S., McLure, R.J., Kukula, M.J., Baum, S.A., O'Dea, C.P., \& Hughes, 
D.H. 2003, \mnras, 340, 1095
\bibitem[Elias et al.(1982)]{Eletal82}
Elias, J.H., Frogel, J.A., Matthews, K., \& Neugebauer, G. 1982, \aj , 87, 1029
\bibitem[Elston et al.(1994)]{Eletal94}
Elston, R., Thompson, K.L., \& Hill, G.J. 1994, \nat , 367, 250
\bibitem[Fan et a.(2000)]{Fanetal00}
Fan, X., et al. 2000, \aj, 119, 1
\bibitem[Fan et al.(2002)]{Fanetal02}
Fan, X., Narayanan, V.K., Strauss, M.A., White, R.L., Becker, R.H., 
Pentericci, L., \& Rix, H.-W. 2002, \aj, 123, 1247
\bibitem[Ferland et al.(1996)]{Feetal96}
Ferland, G.J., Baldwin, J.A., Korista, K.T., et al. 1996, \apj , 461, 683 
\bibitem[Ferrarese \& Merritt(2001)]{FeMe01}
Ferrarese, L. \& Merritt, D. 2001, \apj, 555, L79
\bibitem[Francis et al.(1991)]{Fraetal91}
Francis, P.J., Hewett, P.C., Foltz, C.B., et al. 1991, \apj , 373, 465
\bibitem[Freedman et al.(2001)]{Freetal01}
Freedman, W.L., Madore, B.F., Gibson, B.K., et al. 2001, \apj , 553, 47
\bibitem[Freudling et al.(2003)]{Freu03}
Freudling, W., Corbin, M.R., \& Korista, K.T. 2003, \apj, 587, L67
\bibitem[Fria\c{c}a \& Terlevich(1998)]{FrTe98}
Fria\c{c}a, A.C.S. \& Terlevich, R.J. 1998, \mnras, 298, 399
\bibitem[Gebhardt et al.(2000)]{Gebetal00}
Gebhardt, K., et al. 2000, \apj, 543, L5
\bibitem[Gnedin \& Ostriker(1997)]{GnOs97}
Gnedin, N.Y. \& Ostriker, J.P. 1997, \apj, 486, 581
\bibitem[Granato et al.(2001)]{Gretal01}
Granato, G.L., Silva, L., Monaco, P., Panuzzo, P., Salucci, P., De Zotti, G., 
\& Danese, L. 2001, \mnras, 324, 757
\bibitem[Grandi(1981)]{Gran81}
Grandi, S.A. 1981, \apj , 251, 451
\bibitem[Grandi(1982)]{Gran82}
Grandi, S.A. 1982, \apj , 255, 25 
\bibitem[Green et al.(2001)]{Gretal01}
Green, P.J., Forster, K., \& Kuraszkiewicz, J. 2001, \apj , 556, 727
\bibitem[Grupe et al.(1995)]{Gretal95}
Grupe, D., Beuermann, K., Mannheim, K., Thomas, H.-C., Fink, H.H., \& 
 de Martino, D. 1995, \aap, 300, L21
\bibitem[Guilloteau et al.(1997)]{Guetal97}
Guilloteau, S., Omont, A., McMahon, R.G., Cox, P., \& Petitjean, P. 1997, 
 \aap , 328, L1
\bibitem[Guilloteau et al.(1999)]{Guetal99}
Guilloteau, S., Omont, A., McMahon, R.G., Cox, P., \& Petitjean, P. 1999, 
 \aap , 349, 363
\bibitem[Haiman \& Loeb(1998)]{HaLo98}
Haiman, Z. \& Loeb, A. 1998, \apj, 503, 505
\bibitem[Hamann(1997)]{Hama97}
Hamann, F. 1997, \apjs, 109, 279
\bibitem[Hamann \& Ferland(1992)]{HaFe92}
Hamann, F. \& Ferland, G.J. 1992, \apjl , 381, L53
\bibitem[Hamann \& Ferland(1993)]{HaFe93}
Hamann, F. \& Ferland, G.J. 1993, \apj , 418, 11
\bibitem[Hamann \& Ferland(1999)]{HaFe99}
Hamann, F. \& Ferland, G.J. 1999, \araa , 37, 487
\bibitem[Hill et al.(1993)]{Hietal93}
Hill, G.J., Thompson, K.L., \& Elston, R. 1993, \apj , 414, L1
\bibitem[Horne(1986)]{Horn86}
Horne, K. 1986, \pasp , 98, 609
\bibitem[Iben \& Tutukov(1984)]{IbTu84}
Iben, I. \& Tutukov, A.V. 1984, \apjs , 54, 335
\bibitem[Isaak et al.(1994)]{Isetal94}
Isaak, K.G., McMahon, R.G., Hils, R.E., \& Withington, S. 1994, 
\mnras, 269, L28
\bibitem[Iwamuro et al.\,(2002)]{Iwetal02}
Iwamuro, F., et al. 2002, \apj, 565, 63
\bibitem[Joly(1987)]{Joly87}
Joly, M. 1987, \aap, 184, 33
\bibitem[Kauffmann \& Haehnelt(2000)]{KaHa00}
Kauffmann, G. \& Haehnelt, M.G. 2000, \mnras, 311, 576
\bibitem[Kawara et al.(1996)]{Kaetal96}
Kawara, K., Murayama, T., Taniguchi, Y., \& Arimoto, N. 1996, \apjl , 470, L85
\bibitem[Korista et al.(1996)]{Koetal96}
Korista, K.T., Hamann, F., Ferguson, J., \& Ferland, G.J. 1996, \apj , 461, 641
\bibitem[Kormendy \& Richstone(1995)]{KoRi95}
Kormendy, J. \& Richstone, D. 1995, \araa, 33, 581
\bibitem[Kukula et al.(2001)]{Kuetal01}
Kukula, M.J., et al. 2001, \mnras, 326, 1533
\bibitem[Kurucz(1992)]{Kuru92}
Kurucz, R.L. 1992, in The Stellar Populations of Galaxies, IAU Symp.\,\# 149,
eds.\,B.\,Barbury, A.\,Renzini, Kluwer Dordrecht, p.225
\bibitem[Lanzetta et al.(2002)]{Lanetal02}
Lanzetta, K.M., Yahata, N., Pascarelle, S., Chen, H.-W., 
Fernandez-Soto, A. 2002, \apj, 570, 492
\bibitem[Laor et al.(1995)]{Laoetal95}
Laor, A., Bahcall, J.N., Jannuzi, B.T., Schneider, D.P., \& Green, R.F. 1995,
 \apjs, 99, 1
\bibitem[Laor et al.(1997)]{Laetal97}
Laor, A., Jannuzi, B.T., Green, R.F., \& Boroson, T.A. 1997, \apj , 489, 656
\bibitem[Lilly et al.(1996)]{Liletal96}
Lilly, S.J., Le F\`evre, O., Hammer, F., \& Crampton, D. 1996, \apj, 460, L1
\bibitem[Mathur(2000)]{Mathur00}
Mathur, S. 2000, \mnras, 314, L17
\bibitem[Matteucci \& Greggio(1986)]{MaGr86}
Matteucci, F. \& Greggio, L. 1986, \aap , 154, 279
\bibitem[Matteucci \& Padovani(1993)]{MaPa93}
Matteucci, F. \& Padovani, P. 1993, \apj , 419, 485
\bibitem[Matteucci(1994)]{Matt94}
Matteucci, F. 1994, \aap, 288, 47
\bibitem[Matteucci \& Recchi(2001)]{MaRe01}
Matteucci, F. \& Recchi, S. 2001, \apj, 558, 351 
\bibitem[McIntosh et al.(1999)]{Mcetal99}
McIntosh, D.H., Rieke, M.J., Rix, H.-W., Foltz, C.B., \& Weymann, R.J. 1999,
\apj , 514, 40
\bibitem[Merritt \& Ferrarese(2001)]{MeFe01}
Merritt, D. \& Ferrarese, L. 2001, \apj, 547, 140
\bibitem[Murayama et al.(1998)]{Muetal98}
Murayama, T., Taniguchi, Y., Evans, A.S., et al. 1998, \aj , 115, 2237
\bibitem[Murayama et al.(1999)]{Muetal99}
Murayama, T., Taniguchi, Y., Evans, A.S., et al. 1999, \aj , 117, 1645
\bibitem[Netzer \& Laor(1993)]{NeLa93}
Netzer, H. \& Laor, A. 1993, \apj , 404, L51
\bibitem[Netzer et al.(1985)]{Neetal85}
Netzer, H., Wamsteker, W., Wills, B.J., \& Wills, D. 1985, \apj , 292, 143
\bibitem[Netzer \& Wills(1983)]{NeWi83}
Netzer, H. \& Wills, B.J. 1983, \apj , 275, 445
\bibitem[Netzer et al.(1992)]{NeLaGo92}
Netzer, H., Laor, A., \& Gondhalekar, P.M. 1992, \mnras, 254, 15
\bibitem[Nolan et al.(2001)]{Noetal01}
Nolan, L.A., et al. 2001, \mnras, 323, 308
\bibitem[Ohta et al.(1996)]{Ohetal96} 
Ohta, K., Yamada, T., Nakanishi, K., Kohno, K., Akiyama, M., \& Kawabe, R. 
 1996, \nat, 382, 426
\bibitem[Omont et al.(1996)]{Ometal96} 
Omont, A., Petitjean, P., Guilloteau, S., McMahon, R.G., Solomon, P.M., 
\& Pecontal, E. 1996, \nat, 382, 428
\bibitem[Omont et al.(2001)]{Ometal01}
Omont, A., Cox, P., Bertoldi, F., McMahon, R.G., Carilli, C.L., \& Isaak, K.G.
 2001, \aap , 374, 371
\bibitem[Onken \& Peterson(2002)]{OnPe02} 
Onken, C.A. \& Peterson, B.M. 2002, \apj, 572, 746
\bibitem[Osmer \& Shields(1999)]{OsSh99}
Osmer, P.S. \& Shields, J.C. 1999, in `Quasars and Cosmology',
ASP\,Conf.Ser.\,162, eds. G.\,Ferland \& J.A.\,Baldwin, p.235
\bibitem[Papadopoulos et al.(2000)]{Papaetal00}
Papadopoulos, P.P., et al. 2000, \apj, 528, 626
\bibitem[Perryman et al.(1998)]{Peetal97}
Perryman, M.A.C., Lindegren, L., Kovalevsky, J., et al. 1997, \aap , 323, L49
\bibitem[Peterson \& Wandel(1999)]{PeWa99} 
Peterson, B.M. \& Wandel, A. 1999, \apjl, 521, L95
\bibitem[Peterson \& Wandel(2000)]{PeWa00} 
Peterson, B.M. \& Wandel, A. 2000, \apj, 540, L13
\bibitem[Pettini(1999)]{pet99} Pettini, M. 1999, in Proc.of ESO Workshop 'Chemical Evolution from Zero to High Redshift', ed.\,J.\,Walsh \& M.\,Rosa, p.233
\bibitem[Pogge(2000)]{Pogge00}
Pogge, R.W. 2000, New Astronomy Reviews, 44, 381
\bibitem[Romano et al.(2002)]{Roetal02}
Romano, D., Silva, L., Matteucci, F., \& Danese, L. 2000, \mnras, 334, 444
\bibitem[Sanders et al.(1989)]{Saetal89}
Sanders, D.B., Phinney, E.S., Neugebauer, G., Soifer, B.T., \& Matthews, K.
 1989, \apj , 347, 29
\bibitem[Schneider et al.(1989)]{Schetal89}
Schneider, D.P., Schmidt, M., \& Gunn, J.E. 1989, \aj, 98, 1951 
\bibitem[Shang et al.(2003)]{Shetal03}
Shang, Z., et al. 2003, \apj, 586, 52
\bibitem[Sigut \& Pradhan(1998)]{SiPr98}
Sigut, T.A.A. \& Pradhan, A.K. 1998, \apjl , 499, L139
\bibitem[Sigut \& Pradhan(2003)]{SiPr03}
Sigut, T.A.A. \& Pradhan, A.K. 2003, \apjs, 145, 15
\bibitem[Smecker-Hane \& Wyse(1992)]{SmWy92}
Smecker-Hane, T.A., \& Wyse, R.F.G. 1992, \aj , 103, 1621
\bibitem[Steidel et al.(1999)]{Steetal99}
Steidel, C.C., Adelberger, K.L., Giavalisco, M., Dickinson, M., \&
Pettini, M. 1999, \apj, 519, 1
\bibitem[Storey \& Hummer(1995)]{StHu95}
Storey, P.J. \& Hummer, D.G. 1995, \mnras , 272, 41
\bibitem[Taniguchi et al.(1997)]{Taetal97}
Taniguchi, Y., Murayama, T., Kawara, K., \& Arimoto, N. 1997, \pasj , 49, 419
\bibitem[Thompson et al.(1999)]{Thetal99}
Thompson, K.L., Hill, G.J., \& Elston, R. 1999, \apj , 515, 487
\bibitem[Tinsley(1979)]{Tins79}
Tinsley, B.M. 1979, \apj , 229, 1046
\bibitem[Tremaine et al.(2002)]{Treetal02}
Tremaine, S., et al. 2002, \apj, 574, 740
\bibitem[Tresse \& Maddox(1998)]{TrMa98}
Tresse, L. \& Maddox, S.J. 1998, \apj, 495, 691
\bibitem[van Breugel et al.(1998)]{vanBretal98}
van Breugel, W.J.M., Stanford, S.A., Spinrad, H., Stern, D., \& Graham, J.R.
 1998, \apj, 502, 614
\bibitem[Verner et al.(1999)]{Veetal99}
Verner, E.M., Verner D.A., Korista, K.T., et al. 1999, \apjs , 120, 101
\bibitem[Vestergaard \& Wilkes(2001)]{VeWi01}
Vestergaard, M. \& Wilkes, B.J. 2001, \apjs , 134, 1
\bibitem[Wampler \& Oke(1967)]{WaOk67}
Wampler, E.J. \& Oke, J.B. 1967, \apj , 148, 695
\bibitem[Wandel(1999)]{Wand99}
Wandel, A. 1999, \apj, 527, 649
\bibitem[Warner et al.(2002)]{Waetal02}
Warner, C., et al. 2002, \apj, 567, 68
\bibitem[Weymann et al.(1991)]{Weyetal91}
Weymann, R.J., Morris, S.L., Foltz, C.B., \& Hewett, P.C. 1991, \apj, 373, 23 
\bibitem[Wheeler et al.(1989)]{WST89}
Wheeler, J.C., Sneden, C., \& Truran, J.W. 1989, \araa , 27, 279
\bibitem[Willott et al.(2000)]{Wietal00}
Willott, C.J., Rawlings, S., \& Jarvis, M.J. 2000, \mnras , 313, 237
\bibitem[Wills et al.(1980)]{Wietal80}
Wills, B.J., Netzer, H., Uomoto, A.K., \& Wills, D. 1980, \apj , 237, 319 
\bibitem[Wills et al.(1985)]{WNW85}
Wills, B.J., Netzer, H., \& Wills, D. 1985, \apj , 288, 94 
\bibitem[Yoshii et al.(1996)]{Yoetal96}
Yoshii, Y., Tsujimoto, T., \& Nomoto, K. 1996, \apj , 462, 266
\bibitem[Yoshii et al.(1998)]{Yoetal98}
Yoshii, Y., Tsujimoto, T., \& Kawara, K. 1998, \apjl , 507, L113
\end{thebibliography}
\end{document}